\documentclass[reprint,twocolumn,amsmath,amssymb,aps]{revtex4-1}
\usepackage{graphicx}
\usepackage{bm}
\usepackage{epsfig}
\usepackage{amssymb}
\usepackage{amsfonts}
\usepackage{braket}
\usepackage{color}
\usepackage{epstopdf}
\usepackage{hyperref}
\usepackage{appendix}
\usepackage{float}
\restylefloat{table}
\usepackage{bibentry}
\usepackage{multirow}
\usepackage[caption=false]{subfig}
\usepackage{braket}

\newcommand{\ba}{\begin{eqnarray}}
\newcommand{\ea}{\end{eqnarray}}
\newcommand{\la}{\langle}
\newcommand{\ra}{\rangle}
\newcommand{\bd}{\begin{displaymath}}
\newcommand{\bpm}{\begin{pmatrix}}
\newcommand{\epm}{\end{pmatrix}}
\renewcommand{\v}[1]{{\bf #1}}
\newcommand{\nn}{\nonumber \\}
\newcommand{\bwt}{\begin{widetext}}
\newcommand{\ewt}{\end{widetext}}
\DeclareGraphicsExtensions{.pdf,.png,.jpg}
\graphicspath{{C:\Users\Home\Dropbox\Lab\AKLT}}% Put all figures in this directory.

\usepackage[normalem]{ulem}

\begin{document}
\title{Path Integral for Spin-1 Chain in the Fluctuating Matrix Product State Basis}

\author{Jintae Kim}
\affiliation{Department of Physics, Sungkyunkwan University, Suwon 16419, Korea}
\author{Rajarshi Pal}
\affiliation{Department of Physics, Sungkyunkwan University, Suwon 16419, Korea}
\author{Jin-Hong Park}
\affiliation{Department of Physics, Sungkyunkwan University, Suwon 16419, Korea}
\author{Jung Hoon Han}
\email[Electronic address:$~~$]{hanjemme@gmail.com}
\affiliation{Department of Physics, Sungkyunkwan University, Suwon 16419, Korea}
\date{\today}
\begin{abstract}  
A new method of writing down the path integral for spin-1 Heisenberg antiferromagnetic chain is introduced. In place of the conventional coherent state basis that leads to the non-linear $\sigma$-model, we use a new basis called the fluctuating matrix product states (fMPS) which embodies inter-site entanglement from the outset. It forms an overcomplete set spanning the entire Hilbert space of the spin-1 chain. Saddle-point analysis performed for the bilinear-biquadratic spin model predicts Affleck-Kennedy-Lieb-Tasaki (AKLT) state as the ground state in the vicinity of the AKLT Hamiltonian. Quadratic effective action derived by gradient expansion around the saddle point is free from constraints that plagued the non-linear $\sigma$-model and exactly solvable. The obtained excitation modes agree precisely with the single-mode approximation result for the AKLT Hamiltonian. Excitation spectra for other BLBQ Hamiltonians are obtained as well by diagonalizing the quadratic action.  
\end{abstract}
\maketitle
The coherent state representation of spins is a vital component in constructing path integral of the spin Hamiltonian. It is defined as the eigenstate satisfying $(\v S \cdot \v n ) |\v n \rangle = |\v n \rangle$, where $\v n$ refers to the classical spin orientation and $\v S$ is the spin operator. For multi-spin problems the coherent state basis becomes a direct product $\prod_i \otimes |\v n_i \rangle = |\v n_1 \rangle \otimes |\v n_2 \rangle \otimes \cdots$ over all the sites of the lattice, $i$. Such basis offers an intuitive mapping of the spin Hamiltonian to the path integral form, accomplished by replacing each spin operator by its classical counterpart $\v S_i \rightarrow S \v n_i$ ($S$=spin size). The Berry phase action arises naturally in the coherent state representation as $\sum_i \langle \partial_t \v n_i | \v n_i \rangle  =- i S \sum_i (1-\cos \theta_i) \partial_t \phi_i$, as the sum over all spin variables, in the spherical coordinates $\v n_i= (\sin \theta_i \cos \phi_i, \sin \theta_i \sin \phi_i, \cos \theta_i)$~\cite{Auerbach}.  Haldane argued that this Berry phase leads to a topological action which can critically affect the spin dynamics depending on the parity of  the integer $2S$~\cite{Haldane83}. The essence of the spin-1 antiferromagnetic chain problem is captured in a crisp manner by the Affleck-Kennedy-Lieb-Tasaki (AKLT) model Hamiltonian and its exact ground state~\cite{AKLT87,AKLT88}. Later work showed how to cast the AKLT state as a matrix product state (MPS)~\cite{Klumper92}. More recent developments, mostly taking on the nature of heavy numerics, are on the investigation of dynamics of integer spin chain by enlarging the MPS scheme to encompass low-lying excited states~\cite{verstrate11, Vers13, verstraete13}. 

We present a distinctly new path integral approach to the integer spin chain model. The idea is to employ, instead of the product of site-based coherent states of spins, an entangled, MPS-type basis and develop path integrals therein. This is accomplished by generalizing the MPS formalism to include fluctuating correlated states of spins we call the fluctuating MPS (fMPS). The fMPS states are proven to span the entire Hilbert space of the spin-1 chain and satisfy the completeness relation. Using such fMPS basis, path integral for the spin-1 bilinear-biquadratic (BLBQ) model is constructed following Feynman's canonical prescription. We begin by introducing a singlet bond operator ${\cal S}_{i}^\dag$ and three triplet bond operators $({\cal T}^{x}_{i})^\dag$, $({\cal T}^{y}_{i})^\dag$, $({\cal T}^{z}_{i})^\dag$ defined on a pair of adjacent sites $(i, i+1)$ in the Schwinger boson (SB) representation:
\ba
{\cal S}_{i}^\dag &=& {1 \over \sqrt{2}}(a_i^\dag b_{i+1}^\dag - b_i^\dag a_{i+1}^\dag)\nn
({\cal T}^x_{i})^\dag &=& {1 \over \sqrt{2}}(a_i^\dag a_{i+1}^\dag - b_i^\dag b_{i+1}^\dag)\nn
({\cal T}^y_{i})^\dag &=& {i \over \sqrt{2}}(a_i^\dag a_{i+1}^\dag + b_i^\dag b_{i+1}^\dag)\nn
({\cal T}^z_{i})^\dag &=& { 1 \over \sqrt{2} } ( a_i^\dag b_{i+1}^\dag + b_i^\dag a_{i+1}^\dag).
\label{eq:bond-basis}
\ea
Each boson $a^\dag_i$ and $b^\dag_i$ creates a spin-1/2 particle of up and down orientations, respectively, at the site $i$. 
Notations for the basis operators ${\cal S}_{i}^\dag$, $({\cal T}^{x}_{i})^\dag$, $({\cal T}^{y}_{i})^\dag$, and $({\cal T}^{z}_{i})^\dag$ will be used interchangeably with $(N_{i}^1, N_{i}^2, N_{i}^3, N_{i}^4 )$, respectively. The celebrated AKLT ground state is given simply by the product of singlet bond operators $\prod_{i} {\cal S}_{i}^\dag$ acting on the SB vacuum $|v\ra$. In fact, every bond-product state of the form $N^{\alpha_1}_{1}N^{\alpha_2}_{2}\cdots N^{\alpha_N}_{N}|v\rangle$, with $\alpha_i$'s taking one of the four possibilities in Eq. (\ref{eq:bond-basis}) on a closed chain of length $N$, represents a viable many-body $S=1$ spin state. They also span the entire $3^N$-dimensional space of the spin-1 chain~\cite{KPH19}. 

The SB formalism suggests a way to conveniently express excited states of the spin-1 chain, by writing each bond state as a superposition of the four bond operators introduced in Eq. (\ref{eq:bond-basis}). To be concrete, the bond operator $N_i$ (without the upper index) over the $(i, i+1)$ bond as well as the overall many-body state $|\v N\rangle$ can be introduced as 
\ba
N_i = \sum_{\alpha =1}^4 z_i^{\alpha} N_i^{\alpha} , ~~~~
|\v N\rangle = \left( \prod_{i}N_{i} \right) |v\rangle .\label{eq:bond-coherent-state}
\ea 
One can normalize the complex-valued coefficients $z_i^{\alpha_i }$ according to $\sum_{\alpha_i = 1}^4 z^{\alpha_i}_i \overline{z}^{\alpha_i}_i = 1 ~(\forall i)$, where $\overline{z}^{\alpha_i}_i$ is the complex conjugate of $z^{\alpha_i}_i$. In analogy to the coherent state of spins, we call the above as the bond coherent state. Our goal is to develop a path integral theory of the spin chain within the framework of bond coherent states given in Eq. (\ref{eq:bond-coherent-state}). In particular we want to focus on the spin-1 BLBQ model that contains both the Heisenberg and the AKLT Hamiltonians as special cases. It is given by $H_{\text{BLBQ}} = \sum_{ i } H_i$ where each $H_i$ is \cite{Uimin70, Lai74, Sutherland75}
\ba H_i  & = &{\v S}_{i} \cdot {\v S}_{i+1}+\tan \tau~ ({\v S}_{i} \cdot {\v S}_{i+1})^2\nn
& = & 1 + \tan \tau -(1 + 2\tan \tau)  {\cal{S}}_{i}^\dag {\cal{S}}_{i}+\tan \tau ({\cal{S}}_{i}^\dag {\cal{S}}_{i})^2 . 
\label{eq:Heisen-bond-operator}
\ea
The bond singlet operator ${\cal S}^\dag_i$ [Eq. (\ref{eq:bond-basis})] is used in the second equality. The Heisenberg and the AKLT Hamiltonians are found at $\tau = 0$ and $\tau_0 \equiv \tan^{-1}(1/3)$, respectively. 

An essential ingredient in the path integral construction is the existence of a complete set of continuously varying states $|\v N\rangle$ satisfying the completeness relation $\int [\cal D {\bf{N}} ]  | \bf{N} \ra \la \bf{N}| \propto I$ over a suitable integration measure $[{ \cal D}  {\bf N} ] $. In the Supplementary Information (SI) we present proof that $|\v N\rangle$ defined in Eq. (\ref{eq:bond-coherent-state}) provides such a complete set, over the space of complex-valued coefficients satisfing the constraint $\sum_\alpha |z_i^\alpha |^2 = 1$. An appropriate integration measure for such CP$^3$ fields can be found as~\cite{CP3}: 
\ba
z^1 &=& \cos \chi_i \cos \xi  , \nn
z^2 & = &  e^{i \varphi^x} \cos \chi \sin \xi , \nn
z^3 & = & e^{i \varphi^y} \sin \chi_i \cos \eta, \nn
z^4 & = & e^{i \varphi^z} \sin \chi \sin \eta. 
\label{eq:convention-CP3}
\ea
Here the ranges of angles are $\varphi^\alpha  ~ (\alpha = x, y, z) \in [0, 2\pi]$, $\chi \in [0, {\pi \over 2}]$, $\xi \in [0, 2\pi]$, and $\eta \in [0, 2\pi]$. The integration measure for the $i$-th bond variables is
\ba
\int d\Omega_i & = & \frac{1}{8\pi^5}\int_0^{2\pi}\!\!  d\varphi_i^x \int_0^{2\pi}\!\! d\varphi_i^y \int_0^{2\pi}\!\! d\varphi_i^z\nn
&&\int_0^{2\pi}\!\! d\xi_i\int_0^{2\pi}\!\!  d\eta_i \int^{\pi/2}_{0} d\chi_i \sin 2\chi_i .  \label{eq:measure-CP3}
\ea
Denoting $[{\cal D} \v N ] = \prod_i d \Omega_i$, the desired completeness relation follows as
\ba
\int  [  {\cal D } \v N ] |\v N \rangle \langle \v N | = I_{3^N\times3^N} . 
\ea
Details of the proof are in SI. 

Having found a complete set in the fMPS basis as provided by the bond coherent states $|\v N\rangle$, we can follow Feynman's prescription in constructing the path integral by evaluating the time evolution amplitude over an infinitesimal time interval $\Delta t$: 
\ba 
&&{ \la \v N (t + \Delta t ) | e^{- i  \Delta t H} | \v N (t) \ra  \over \la \v N(t) | \v N (t) \ra   } \simeq \nn
&&  \exp \left( i \Delta t [-i {\la \partial_t \v N (t) | \v N(t) \ra \over \la \v N(t) | \v N (t) \ra} - {\la \v N (t)| H | \v N (t) \ra \over \la \v N(t) | \v N (t) \ra } ] \right),  \label{eq:path-overlap} 
\ea
with $|{\bf N}(t)\ra $ denoting the bond coherent state [Eq. (\ref{eq:bond-coherent-state})] at time $t$. Each term in the action
requires evaluation of the overlap of one many-body state $|\v N\rangle$ with another state, {\it e.g.} $|\partial_t \v N\rangle$, $H |\v N\rangle$. In general this is a formidable problem, circumvented in the usual path integral approach only by use of the product state basis in which the inter-site correlations are absent. The employment of product basis states implies that intricate correlations inherent in the model remain ``hidden" in the action, demanding a lot of analysis of the resulting action to uncover them. By introducing correlated basis from the start, as we do with the fMPS basis, one can hope that much of the correlations in the model has already been built in, resulting in the effective action that is simple to analyze. Such seems to be the case with the spin-1 chain problem. 

The process begins by identifying the lowest-energy configuration in the variational space of fMPS. The spin-1 BLBQ Hamiltonian [Eq. (\ref{eq:Heisen-bond-operator})] has the expectation value
\ba
E = \sum_i { \langle \v N |H_i|\v N \ra  \over \langle \v N | \v N \rangle }  
\ea
for some fMPS state $|\v N\rangle$. Anticipating more or less uniform variational state to give the lowest energy, we first search the space of fMPS where each bond operator $N_i = \sum_\alpha z^\alpha N^\alpha_i $ is uniform (site-independent). One can further invoke rotational symmetry within the triplet space to confine the search only to the $(N^1, N^4)$ sector, parameterized by $z^1 = \cos \frac{\theta}{2},~z^4 = e^{i\phi} \sin \frac{\theta}{2}$. For this class of uniform MPS we find $z^1 = 1$ gives the lowest energy $E=-4/3$. This is nothing but the AKLT state in the MPS form. Our variational search is limited to the range of parameters in the BLBQ model $-\pi/4 < \tau < \pi/4$ where the ground states are known to be gapped and paramagnetic~\cite{Uimin70, Lai74, Sutherland75}.

\begin{figure}
\includegraphics[width=0.46\textwidth]{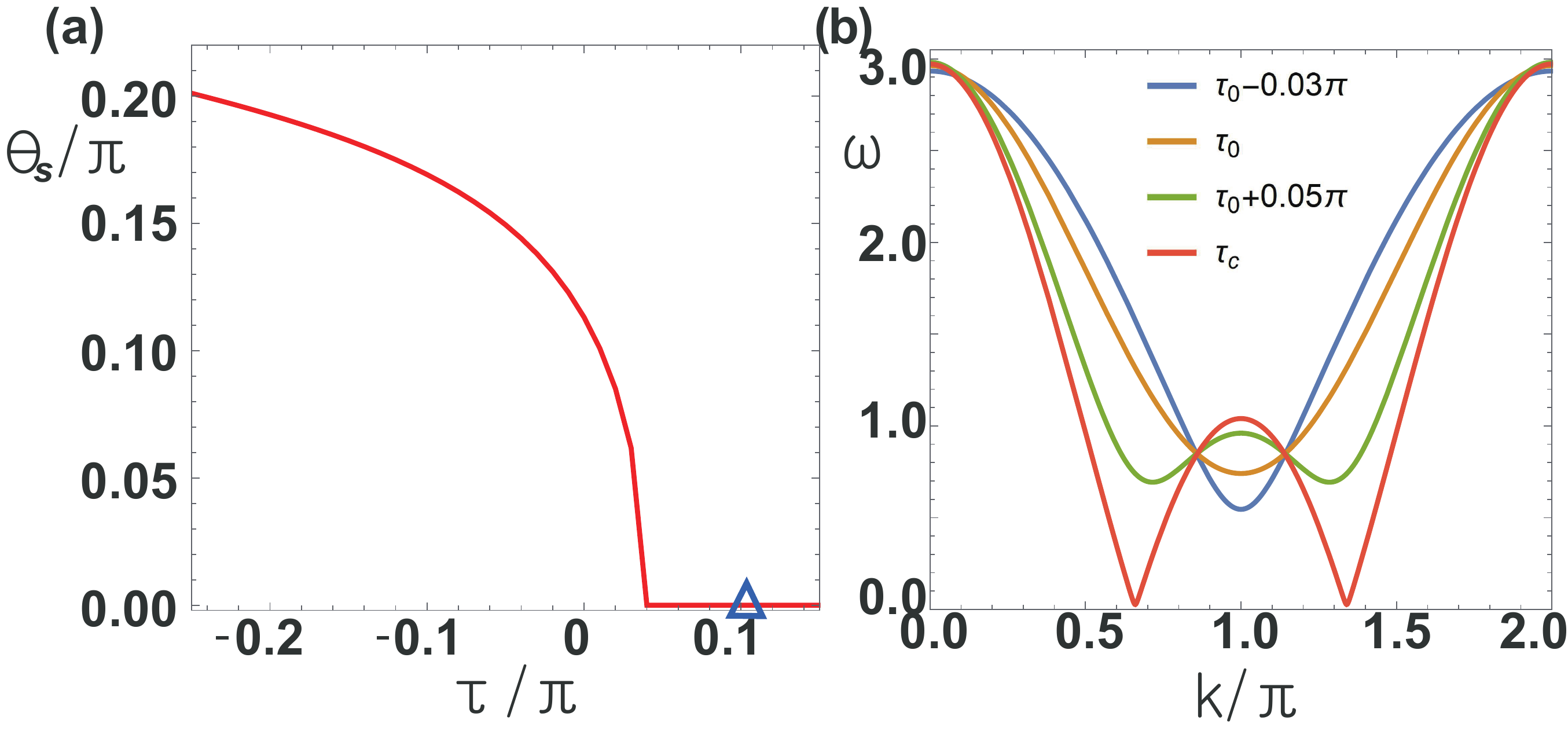} 
\caption{(a) Mixing angle $\theta_s$ at the saddle point of the energy vs. the angle $\tau$ in the BLBQ Hamiltonian. $\theta_s = 0$ is the AKLT state. The AKLT Hamiltonian is found at $\tau_0 =\tan^{-1} \frac{1}{3}$, indicated by the blue triangle. (b) Dispersion relation obtained from the effective action at various values of $\tau$. An adjustment $k \rightarrow k + \pi$ was made in the formula, Eq. (\ref{eq:omega}), to have our plot agree with other plots of the excitation spectra in the literature. The energy minimum occurs away from $k=\pi$ as $\tau >\tau_0$ and becomes zero at $\tau=\tau_c$ (see text for details). }
\label{fig:3}
\end{figure}

The other type of fMPS ansatz investigated assumes uniform singlet, but staggered triplet configuration according to $N_i = z^1 N^1 + (-1)^i z^4 N^4$. The triplet amplitude alternates in sign from bond to bond. In this case, the energy-minimizing state is found at $z^2 = {\rm real}$, with the finite mixing angle $\theta_s$ as shown in Fig. \ref{fig:3}(a). This is a symmetry-breaking state, as evidenced by explicit calculations showing $\braket{S_i^x}=\braket{S_i^y}=0$ but $\braket{S_i^z}=-\braket{S_{i+1}^z} \neq 0$. A plot of $\langle S^z_i \rangle$ for the staggered variational MPS state is presented in SI for completeness. In fact one can prove $|S^z_i | = 1$ at the mixing angle $\theta=\pi/2$.  The appearance of magnetic ground state is an artifact of the variational calculation and runs counter to the well-known result that only paramagnetic ground states exist for the BLBQ Hamiltonian for $- \pi/4 <   \tau  < \pi/4$.  To make further analysis possible, we will henceforth confine our attention to $0.04\pi \lesssim \tau <\pi/4$ for which the variational minimum is indeed found at $\theta_s =0$ (AKLT state); see Fig. \ref{fig:3}(a). 

Effective action can be derived by computing the various overlaps in Eq. (\ref{eq:path-overlap}) for small fluctuations around the AKLT saddle point. In general this is a highly nontrivial task, as overlaps of many-body states are involved in the computation. Thanks to the well-known gapped spectrum of the AKLT-type ground state, however, we may reasonably anticipate that fluctuations are dominated by the creation of broken singlet bonds, or solitons. Here the solitons are the three triplet bonds [Eq. (\ref{eq:bond-basis})] taking the place of the singlet bond in the AKLT state. Assuming a low-energy manifold dominated by 0, 1, or 2 solitons in the whole chain, we express the general fMPS states as an expansion in the number of solitons,
\ba |\v N\rangle & \approx & \prod_{i} \Bigl( N_i^1 + \sum_{\alpha=2}^4 (-1)^i z_i^\alpha N_i^\alpha \Bigl) |v\rangle \nn
& \approx & \ket{A}+\sum_i\ket{\mathcal{T}_i}+\sum_{j<i}\ket{\mathcal{T}_j \mathcal{T}_i}+\cdots  . \label{eq:staggered-form}
\ea
The staggered sign $(-1)^i$ introduced in the first line allows a smooth expression of the effective action, without the alternating sign. 
Smallness of the triplet amplitudes $|z_i^{\alpha}|\ll1$ assumes that we are expanding the action around the AKLT saddle point denoted by $\ket{A}$. One-soliton state is written as $\ket{\mathcal{T}_i}$ and given by replacing the singlet bond operator $\mathcal{S}_{i}^{\dagger}$ in the AKLT state by triplet creation operator $\mathcal{T}_i^{\dagger}=\sum_{\alpha=2}^4(-1)^i z_i^{\alpha}N_i^{\alpha}$. 
Similarly, $\ket{\mathcal{T}_i\mathcal{T}_j}$ is obtained by introducing a pair of triplet creation operators at $(i,i+1)$ and $(j,j+1)$ bonds. Effective action can be derived by evaluating the overlaps in the path integral (\ref{eq:path-overlap}) systematically up to second power in $z_i^\alpha$'s. For consistent implementation of the staggered bond factor $(-1)^i$ on a closed chain we adopt even $N$ for the size of the chain.  

With abbreviations $\eta \equiv 1/3$ and $f_{ij} = \eta^{|i-j|} + \eta^{N- |i-j|}$, one can prove (see SI for computational details)  
\ba
\braket{\v N|\v N} &=& \left(\frac{3}{2}\right)^N\Bigl[1 \! +\!  \sum_{i, \alpha}  | z_{i}^{\alpha} |^2 \! +\!  2 \sum_{i \neq j , \alpha} f_{ij} x_{i}^{\alpha} x_{j}^{\alpha}  \Bigl]\nn
\braket{\partial_t \v N|\v N} &=& \left(\frac{3}{2}\right)^N\sum_{i,j, \alpha} f_{ij} ( \partial_t {\overline{z}}_{i}^{\alpha} )  z_{j}^{\alpha} . 
\ea
We break up the complex coefficients $z^\alpha_i = x^\alpha_i + i y^\alpha_i$ as real and imaginary parts. 
The ``Berry phase" term follows as
\ba
-i\frac{\braket{\partial_t \v N|\v N}}{\braket{\v N|\v N}} &=& -i \sum_{i,j, \alpha }f_{ij}  ( \partial_t {\overline{z}}_{i}^{\alpha} )  z_{j}^{\alpha} .  \label{eq:newBerry}
\ea
Different spin orientations $\alpha$ do not mix in the effective action in observance of the rotational symmetry in the space of triplet excitations. Note that $f_{ij} \neq 0$ for $i\neq j$ leads to an unusual, long-ranged Berry phase action in distinct comparison to the coherent-state based action involving only the local terms $-i \sum_i ( \partial_t {\overline{z}}_{i} )  z_{i}$. While the $z_i$'s are CP$^1$ fields in the conventional representation of the Berry phase action with constraints $|z_i |^2 = 1$, our fields $z^\alpha_i$ do not have such constraints except that they be small in amplitudes. In other words, $z^\alpha_i$'s are ``free" fields, which make the subsequent calculations easy to handle. 

Energy functional $E_i = \langle \v N | H_i | \v N\rangle/\braket{\v N|\v N}$ can be worked out in the similar approximation scheme (details are in SI):
\bwt
\ba\label{eq:Energy}
E_i & = &  -\frac{4}{3}+2 \tan \tau+ {8 \over 9} (2 -\tan \tau)\sum_{\alpha} | z_{i}^{\alpha} |^2 \nn
&& - {32 \over 3}  \left(1 \! - \!3\tan \tau \right)  \left(  \sum_{j<k<i,i<j<k } \eta^{N-|j-k|}   +   \sum_{j<i<k} \eta^{|j-k|} \right)  \sum_{\alpha} x^\alpha_j x^\alpha_k  . 
\ea
\ewt
Although somewhat lengthy, this is still a quadratic action in terms of free and independent fields $z^\alpha$ and easily diagonalizable. The expression becomes remarkably simple at the AKLT point $\tau_0 = \tan^{-1} (1/3)$ as all the long-ranged interaction terms in the second line vanish. 

Equation of motion follows readily from varying the Berry phase action (\ref{eq:newBerry}) and the total energy $\sum_i E_i$: 
\ba
&&\sum_{j}\bigl( \eta^{|j-i|}+ \eta^{N-|j-i|} \bigr) \dot{x}_{j}^{\beta} -{8 \over 9} ( 2 - \tan \tau)y_{i}^{\beta}=0 , \nn
&&\sum_{j} \bigl( \eta^{|j-i|}+ \eta^{N-|j-i|} \bigr)\dot{y}_{j}^{\beta} +{ 8 \over 9} (2 - \tan \tau)x_{i}^{\beta} \nn
&& ~~~~~~~~~~~~~ = { 8 \over 3} \left(1 - 3 \tan \tau \right) \sum_{j\neq i} F_{i,j}x_j^{\beta} 
\ea
where $F_{i,j}$ is defined as
\ba
F_{i,j} = 2\left((|j\! -\! i|-1)\eta^{|j\!-\!i|} \! +\! (N-|j\!-\!i|-1)\eta^{N-|j-i|}\right).\nonumber 
\ea
Each triplet branch $\alpha=2,~3,~4$ acts independently. The equation of motion in the Fourier space, $x^\alpha_{j}=\sum_{k,\omega}X^\alpha_{k,\omega}e^{i k j-i \omega t}, ~ y_{j}^\alpha=\sum_{k,\omega}Y^\alpha_{k,\omega} e^{i k j-i \omega t}$ gives (omitting spin indices $\alpha$ and $(k, \omega)$)
\ba
&&i\omega G(k) X = - {8 \over 9} (2 - \tan \tau )Y \nn
&&i\omega G(k) Y \!=\! {8 \over 9} \left[2 \!-\! \tan \tau  \!+\!  3 (3\tan \tau \!-\! 1) F(k) \right] X \ea
where $F(k)$, $G(k)$ are
\ba
F(k)&=&\sum_{j\neq i}F_{i,j}e^{ik(j-i)}=\frac{9 \cos 2 k-6 \cos k+1}{(5-3 \cos k)^2}\nn
G(k)&=&\sum_{j}\Bigl( \eta^{|j-m|} \!+ \!\eta^{-|j-m|+N}\Bigl)e^{i k (j-i)}=\frac{4}{5-3\cos k} \nonumber
\ea
after taking the large $N$ limit. After all, $\omega$ becomes
\ba
\omega={8 \bigl[  (2\! -\!  \tan \tau ) (2\!-\! \tan \tau \!+\!3  (3\tan \tau \!-\! 1)F(k) ) \bigr]^{1/2} \over 9 G(k) }  \nn \label{eq:omega}
\ea

Plots of the dispersion for several values of $\tau$ are shown in Fig. \ref{fig:3}(b).
The choice of $\tau$ in plotting the dispersion is necessarily confined to the region where the variational MPS ground state equals the non-magnetic AKLT state, i.e. $\theta_s = 0$ in Fig. 1(a). The well-known single-mode approximation (SMA) for the excitation energy in the AKLT model is perfectly recovered by the above dispersion formula at $\tan \tau = 1/3$: $\omega(k) = (10/27)(5-3 \cos k)$. It differs from the conventional expression $(5/27)(5+ 3\cos k)$ \cite{AAH88} only due to the fact that our definition of AKLT Hamiltonian is twice that of the conventional one, and the origin of momentum has been displaced by $\pi$ due to the staggered factor $(-1)^i$ we used in the gradient expansion, Eq. (\ref{eq:staggered-form}). The dispersion formula derived in Eq. (\ref{eq:omega}) goes beyond the AKLT point and captures the excitation spectrum for a family of BLBQ models. It is also worth noting that the structure factor $G(k)$, which dominates the dispersion at the AKLT point, entirely comes from the structure of the Berry phase action, having nothing to do with the form of the energy functional. 
Our dispersion formula becomes gapless at $\tau=\tan^{-1} \frac{23}{37} \approx 0.177 \pi $ $k=\pi-\cos^{-1}\frac{13}{27} \approx 0.660\pi$  while the actual BLBQ model becomes gapless at $\tau = \pi/4$ and $k=2\pi/3$. Such discrepancy is expected given the simple nature of  our MPS ansatz. Nevertheless it is nontrivial that gapless point occurs in our approach at values that are in fair proximity to the exact values. Low-energy modes of the BLBQ Hamiltonian have been worked out elsewhere using extensive numerical methods, e.g. see Fig. 3 in Ref.~\cite{Vers13}. Minimum of the dispersion occurs away from $k=\pi$ as $\tau$ increases beyond the AKLT value $\tau_0$, in agreement with the behavior exhibited by our dispersion formula, Eq. (\ref{eq:omega}). 

The path integral construction of the spin-1 Hamiltonian using the more general MPS formalism was made in earlier works~\cite{green,hallam}. Our construction makes explicit use of the completeness of the singlet/triplet bond states and gives analytical expression of the low-energy dispersion. A different kind of path integral construction using the squeezed states was recently advanced~\cite{mathey} and applied to study the dynamics of cold atoms. 

\acknowledgments
J. H. H. was supported by Samsung Science and Technology Foundation under Project Number SSTF-BA1701-07.

\appendix

\begin{widetext}
\section{Completeness relation of coherent state}
\begin{figure*}[t]
\centering
\includegraphics[width=160mm]{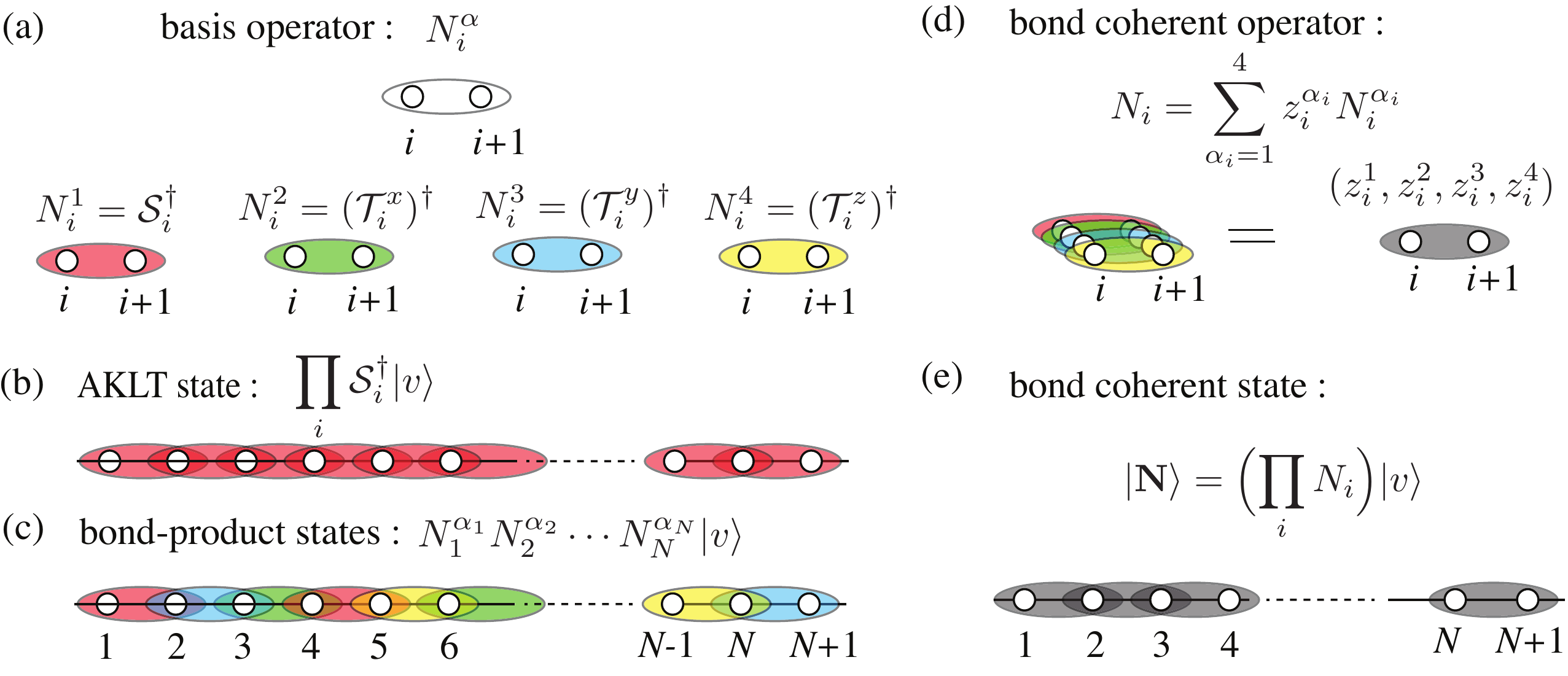}
\caption{Schematic figure of basis operator, AKLT state, bond-product states, bond coherent operator, and bond coherent state. (a) The basis operator $N^\alpha_i$ is introduced with color-coded for each $\alpha$ $(\alpha = 1,2,3,4)$. (b) The AKLT state is demonstrated. (c) The bond-product states is defined in the form of $N^{\alpha_1}_{1} N^{\alpha_2}_{2} \cdots N^{\alpha_N}_{N} | v \ra$ with $\alpha_i$'s taking on one of the four possibilities in Eq. (\ref{eq:bond-basis}). (d) The bond coherent operator is introduced with complex-valued coefficients $(z^1_i, z^2_i, z^3_i, z^4_i)$. Finally, the bond coherent state is constructed in (e).} \label{fig:coherent-states}
\end{figure*}
We offer proof here that the many-body coherent state written in Eq. (2) in the main text does meet this requirement and may well serve as basis for path integral construction. Note that different bond product states (see Fig.~\ref{fig:coherent-states}) 
\ba
\ket{\alpha} = \ket{N^{\alpha_1} N^{\alpha_2} \cdots N^{\alpha_N}}=N^{\alpha_1}_1 N^{\alpha_2}_2 \cdots N^{\alpha_N}_N\ket{v}
\ea
are not mutually orthogonal, i.e. $\langle \alpha' |\alpha \rangle \neq 0$ for $\alpha' \neq \alpha$. This suggests that the states spanned by $|\alpha\rangle$ are over-complete. Nevertheless, one can prove the completeness

\ba
\label{eq:tripovclmp}
\sum_{\{\alpha\}}  |\alpha \ra \la \alpha | =2^{N} I_{3^{N} \times 3^{N}} \label{eq:complete1} 
\ea
where $\sum_{\{\alpha\}}=\sum_{\alpha_1,\alpha_2,\cdots, \alpha_{N}}$ and each $\alpha_i$ runs over the four indices 1 through 4. 

We begin by introducing the four  states  
\begin{align}
|A_1 \ra &= M a_i^\dag a_{i+1}^\dag|v \ra &|A_2 \ra &= M a_i^\dag b_{i+1}^\dag|v \ra \nn
|A_3 \ra &= M b_i^\dag a_{i+1}^\dag|v \ra &|A_4 \ra &= M b_i^\dag b_{i+1}^\dag|v \ra
\end{align}
where $M$ is an arbitrary Schwinger Boson (SB) expression of creation operators such that $|A_1\rangle$ through $|A_4 \rangle$  are valid states of the spin-1 chain. We then form the following combination 
\begin{align}
|C_1 \ra &= \frac{1}{\sqrt{2}}( |A_2 \ra - |A_3 \ra) = M{\cal S}^\dag_{i} |v \ra &|C_2 \ra &= \frac{1}{\sqrt{2}} (|A_1 \ra - |A_4 \ra) = M({\cal T}^{x}_{i})^\dag |v \ra  \nn
|C_3 \ra &= \frac{i}{\sqrt{2}} (|A_1 \ra + |A_4 \ra) = M({\cal T}^{y}_{i})^\dag |v \ra&|C_4 \ra &= \frac{1}{\sqrt{2}} (|A_2 \ra + |A_3 \ra) = M({\cal T}^{z}_{i})^\dag |v \ra, 
\end{align}
which clearly obeys
\ba \label{eq:identity} \sum_{i=1}^4 |C_i \ra \la C_i| = \sum_{i=1}^4 |A_i \ra \la A_i|.  \ea 
The completeness (\ref{eq:complete1}) will be proved by applying  Eq.~(\ref{eq:identity}) repeatedly. Let us start by taking the sum over a single index $\alpha_N$ using Eq.~(\ref{eq:identity}), which gives
\ba
\sum_{ \{ \bm \alpha \} } | \bm \alpha \ra \la \bm \alpha |  =\sum_{\{ \alpha_1, \cdots, \alpha_{N-1} \}}\sum_{s'_N, s_1}  | N^{\alpha_1} N^{\alpha_2}\cdots N^{\alpha_{N-1}} s'_N s_1 \ra \la N^{\alpha_1}N^{\alpha_2}\cdots N^{\alpha_{N-1}} s'_N s_1  | .
\ea
Each $s_i = \pm 1/2$ stands for the two SB operators $a^\dag_i$ and $b^\dag_i$ at site $i$. As one can see from the expression on the right side, we are assuming a periodic chain of length $N$, where the $N$-th site is adjacent to $i=1$ site. The conversion scheme just carried out for the $(N,1)$ bond can be applied for every bond independently, thus giving rise to  
\ba
\sum_{ \{ \bm \alpha \} } | \bm \alpha \ra \la \bm \alpha | &=&\sum_{s'_1,s_1}\sum_{s'_2,s_2}\cdots\sum_{s'_N,s_N}|s'_1 s_2\cdots s'_N s_1 \ra \la s'_1 s_2\cdots s'_N s_1|\nn
&=&\sum_{s'_1,s_1}\sum_{s'_2,s_2}\cdots\sum_{s'_N,s_N}\prod_{i}^N \otimes |s_is_i' \ra \la s_is_i'| .  \label{eq:212} 
\ea
Keep in mind that the spin labels $s_i, s'_i = \pm 1/2$ are those of the Schwinger bosons, and not yet of the physical spins. The conversion to the physical spin degrees of freedom $|S_i \rangle$ satisfying $S_i^z\ket{S_i}=S_i \ket{S_i}$ $(S_i = +1,~0,~-1)$ is done by observing that physical spin states are defined as
\begin{align}
a_i^{\dag}b_i^{\dag}|v \ra = b_i^{\dag}a_i^{\dag}|v \ra&=|0_i \ra &\frac{1}{\sqrt{2}} ( {a_i^{\dag}} )^2|v \ra&= |1_i\ra&\frac{1}{\sqrt{2}}( {b_i^{\dag}} )^2|v \ra&= |{-1}_i\ra,
\end{align}
and that
\ba 
\sum_{s_2,s_2' = \pm 1/2 }|s_2,s_2' \ra \la s_2,s_2'| &=& |aa \ra \la aa| + 2|a b\ra \la a b| + |bb\ra \la b b| \nn
&=& 2( |1 \ra \la 1 | +|0 \ra \la 0 | + |-1 \ra \la -1 |) =  2I_{3 \times 3} . 
\ea
% 
%This proves that Eq. (\ref{eq:212}) is complete after all, and we arrive at the proof of Eq.~(\ref{eq:tripovclmp}) for completeness. 
Denoting ${\cal D } \Omega = \prod_i d \Omega_i$, the desired completeness relation would be  
\ba
\int { \cal D } \Omega \sum_{\{\alpha_N\}}\sum_{\{\beta_N\}} \prod_i \prod_j  z^{\alpha_i}_i \overline{z}^{\beta_j}_j |N^{\alpha_1} \cdots N^{\alpha_N} \rangle \langle N^{\beta_1} \cdots N^{\beta_N}| = I_{3^N\times3^N},\label{eq:single-bond-completeness}
\ea
with  $i,~j$ running  over the lattice sites. One can easily prove the integral remains non-zero only for $i=j$ in the above, therefore
\ba
\eqref{eq:single-bond-completeness} &=& \sum_{\{\alpha_N\}}\sum_{\{\beta_N\}} \prod_i\Bigl( \int d\Omega_i z_i^{\alpha_i} \overline{z}_i^{\beta_i} \Bigl) |N^{\alpha_1} \cdots N^{\alpha_N} \rangle \langle N^{\beta_1} \cdots N^{\beta_N}|.
\ea
We can readily show that $\int d\Omega_i z_i^{\alpha_i} \overline{z}_i^{\beta_i}=\frac{1}{2}\delta_{\alpha_i \beta_i}$, 
concluding the right-hand side of the above expression does match the sum over the bond product states in Eq. (\ref{eq:complete1}). 
Since the completeness of the latter is already proven, we conclude that the bond coherent states under the CP$^3$ parameterization are also complete.

\section{Saddle point solution of the fMPS ansatz}

%\subsection{General overlap calculation between MPS States}

The SB representation of the AKLT ground state $|A\rangle$ allows an equivalent matrix product state (MPS) expression. Schematically, the correspondence can be expressed as
%\cite{zittartz92}
\ba
|A\rangle = [ \prod_i N^1_i ]  |v\rangle \leftrightarrow  \sum_{\{s\} }{\rm Tr} [M_1^{(s_{1})}M_1^{(s_{2})}\cdots M_1^{(s_{N})}] |\{s \} \rangle.\label{eq:AKLT-MPS}
\ea
The summation over all possible spin orientations $\{s\} = \{ s_1, \cdots, s_N \}$ is performed in the second line. The $2\times2$ matrices defined here for each spin orientation $s=1$, $0$, $-1$ are given by
\begin{align}
{M}_1^{(1)}&= \begin{pmatrix}
0 & 1\\
0 & 0\end{pmatrix} &
M_1^{(0)}&=\begin{pmatrix}
-\frac{1}{\sqrt{2}} & 0\\
0 & \frac{1}{\sqrt{2}}\end{pmatrix} &
M_1^{(-1)} &= \begin{pmatrix}
0 & 0\\
-1 & 0\end{pmatrix} .
\end{align}
Such one-to-one correspondence extends well beyond the AKLT state, and in fact covers an {\it arbitrary} state in which one replaces the SB singlet $N_i^1$ by one of the triplets $N_i^\alpha~(\alpha=2,~3,~4)$ defined by Eq.(1) in the main text.

First we introduce an additional set of matrices
\begin{align}\label{eq:sol}
M_{2}^{(1)} &= \begin{pmatrix}
1 & 0\\
0 & 0\end{pmatrix} &
M_{2}^{(0)} &= \begin{pmatrix}
0 & -\frac{1}{\sqrt{2}}\\
\frac{1}{\sqrt{2}} & 0\end{pmatrix} &
M_{2}^{(-1)} &= \begin{pmatrix}
0 & 0\\
0 & -1 \end{pmatrix}\nn
M_{3}^{(1)} &= \begin{pmatrix}
i & 0\\
0 & 0\end{pmatrix}\nonumber&
M_{3}^{(0)} &= \begin{pmatrix}
0 & \frac{i}{\sqrt{2}}\\
\frac{i}{\sqrt{2}} & 0\end{pmatrix} &
M_{3}^{(-1)} &= \begin{pmatrix}
0 & 0\\
0 & i\end{pmatrix}\\
M_{4}^{(1)} &= \begin{pmatrix}
0 & 1\\
0 & 0\end{pmatrix} &
M_{4}^{(0)} &= \begin{pmatrix}
\frac{1}{\sqrt{2}} & 0\\
0 & \frac{1}{\sqrt{2}}\end{pmatrix} &
M_{4}^{(-1)} &= \begin{pmatrix}
0 & 0\\
1 & 0\end{pmatrix} .
\end{align}
The upper index $s$ still refers to the spin orientation of the basis state and the three lower indices $2,~3,~4$ respectively correspond to the $N^2,~N^3,~N^4$. Whenever a particular singlet $N_{i}^1$ in the AKLT state is replaced by the triplet $N_i^\alpha$, one replaces $M_1^{(s_i)}$ by $M_{\alpha}^{(s_i)}$. Therefore an arbitrary bond product state can be expressed in terms of MPS representation as below.
\ba \label{eq:triplon-MPS}
\ket{N^{\alpha_1} N^{\alpha_2} \cdots N^{\alpha_N}} =\sum_{\{s\}}\textrm{Tr}[M_{\alpha_1}^{(s_{1})}M_{\alpha_2}^{(s_{i})}\cdots M_{\alpha_N}^{(s_{N})}]\ket{\{ s\}} .
\ea
%
% A nice way to account for the sign factors in (\ref{eq:AKLT-MPS}) and (\ref{eq:triplon-MPS}) is to remember the replacement rule
% %
% \ba\label{eq:rep-rule}
% [ {\cal S}_{i,i+1} ]^\dag \rightarrow - M^{(s)}_0 ~~~
% [{\cal T}_{i,i+1}^{\alpha}]^{\dagger} \rightarrow M^{(s)}_{\alpha+2}  .
% \ea
%

Now that every SB state has an equivalent MPS representation, their overlaps can also be evaluated by invoking their MPS forms. For two arbitrary MPS states $|\psi\rangle$ and $|\psi'\rangle$, their overlap is
\ba
\ket{\psi} &=& \sum_{\{s\}}\textrm{Tr}[A_1^{(s_{1})}A_2^{(s_{2})}\cdots A_N^{(s_{N})}]\ket{\{s\}}\nn
\ket{\psi'} &=& \sum_{\{s\}}\textrm{Tr}[B_1^{(s_{1})}B_2^{(s_{2})}\cdots B_N^{(s_{N})}]\ket{\{s\}}\nn
\braket{\psi'|\psi} &=& \sum_{\{s\}}\textrm{Tr}[B_1^{(s_{1})}B_2^{(s_{2})}\cdots B_N^{(s_{N})}]\textrm{Tr}[A_1^{(s_{1})}A_2^{(s_{2})}\cdots A_N^{(s_{N})}]
\ea
where $A_i$, $B_i=M_1$, $M_2$, $M_3$, $M_4$. Employing some matrix identities
\ba
\textrm{Tr}[A]\textrm{Tr}[B] &=& \textrm{Tr}[A\otimes B]\nn
ABC\otimes DEF &=& (A\otimes D)(B\otimes E)(C\otimes F)
\ea
one can rewrite the overlap
\ba
\braket{\psi'|\psi} &=&\sum_{\{s\}} \textrm{Tr}[(B_1^{(s_{1})}B_2^{(s_{2})}\cdots B_N^{(s_{N})})\otimes(A_1^{(s_{1})}A_2^{(s_{2})}\cdots A_N^{(s_{N})})]\nn
&=&\sum_{\{s\}}\textrm{Tr}[(B_1^{(s_{1})}\otimes A_1^{(s_{1})})(B_2^{(s_{2})}\otimes A_2^{(s_{2})})\cdots (B_N^{(s_{N})}\otimes A_N^{(s_{N})})]\nn
&=&\textrm{Tr}[(\sum_{s_1}B_1^{(s_{1})}\otimes A_1^{(s_{1})})(\sum_{s_2}B_1^{(s_{2})}\otimes A_1^{(s_{2})})\cdots(\sum_{s_N}B_1^{(s_{N})}\otimes A_1^{(s_{N})})].
\ea
Since there are 4 possibilities for each matrix, there are 16 cases of the direct product $\sum_{s_i}B_i^{(s_{i})}\otimes A_i^{(s_{i})}$ in all. Define a matrix $M_{ij}$ and overlap MPS form of $\braket{\psi'|\psi}$
\ba
\label{eq:mps-overlap}
M_{ij} &=& \sum_{s}M_{i}^{(s)}\otimes M_{j}^{(s)}\nn
\braket{\psi'|\psi} &=& \textrm{Tr}[M_{i_1 j_1} M_{i_2 j_2} \cdots M_{i_N j_N} ]
\ea
The sixteen $M_{ij}$ matrices are given by
\begin{align}
M_{11} &= \begin{pmatrix}\label{eq:M_ij}
\frac{1}{2} & 0 & 0 & 1\\
0 & -\frac{1}{2} & 0 & 0\\
0 & 0 & -\frac{1}{2} & 0\\
1 & 0 & 0 & \frac{1}{2}\end{pmatrix}&
M_{12} &= \begin{pmatrix}
0 & \frac{1}{2} & 1 & 0\\
-\frac{1}{2} & 0 & 0 & 0\\
0 & 0 & 0 & -\frac{1}{2}\\
0 & 1 & \frac{1}{2} & 0\end{pmatrix}&
M_{13} &= \begin{pmatrix}
0 & -\frac{i}{2} & i & 0\\
\frac{-i}{2} & 0 & 0 & 0\\
0 & 0 & 0 & \frac{i}{2}\\
0 & -i & \frac{i}{2} & 0\end{pmatrix}&
M_{14} &= \begin{pmatrix}
-\frac{1}{2} & 0 & 0 & 1\\
0 & -\frac{1}{2} & 0 & 0\\
0 & 0 & \frac{1}{2} & 0\\
-1 & 0 & 0 & \frac{1}{2}\end{pmatrix}\nn
M_{21} &= \begin{pmatrix}
0 & 1 & \frac{1}{2} & 0\\
0 & 0 & 0 & -\frac{1}{2}\\
-\frac{1}{2} & 0 & 0 & 0\\
0 & \frac{1}{2} & 1 & 0\end{pmatrix}&
M_{22} &= \begin{pmatrix}
1 & 0 & 0 & \frac{1}{2}\\
0 & 0 & -\frac{1}{2} & 0\\
0 & -\frac{1}{2} & 0 & 0\\
\frac{1}{2} & 0 & 0 & 1\end{pmatrix}&
M_{23} &= \begin{pmatrix}
i & 0 & 0 & -\frac{i}{2}\\
0 & 0 & -\frac{i}{2} & 0\\
0 & \frac{i}{2} & 0 & 0\\
\frac{i}{2} & 0 & 0 & -i\end{pmatrix}&
M_{24} &= \begin{pmatrix}
0 & 1 & -\frac{1}{2} & 0\\
0 & 0 & 0 & -\frac{1}{2}\\
\frac{1}{2} & 0 & 0 & 0\\
0 & \frac{1}{2} & -1 & 0\end{pmatrix}\nn
M_{31} &= \begin{pmatrix}
0 & i & -\frac{i}{2} & 0\\
0 & 0 & 0 & \frac{i}{2}\\
-\frac{i}{2} & 0 & 0 & 0\\
0 & \frac{i}{2} & -i & 0\end{pmatrix}&
M_{32} &= \begin{pmatrix}
i & 0 & 0 & -\frac{i}{2}\\
0 & 0 & \frac{i}{2} & 0\\
0 & -\frac{i}{2} & 0 & 0\\
\frac{i}{2} & 0 & 0 & -i\end{pmatrix}&
M_{33} &= \begin{pmatrix}
-1 & 0 & 0 & -\frac{1}{2}\\
0 & 0 & -\frac{1}{2} & 0\\
0 & -\frac{1}{2} & 0 & 0\\
-\frac{1}{2} & 0 & 0 & -1\end{pmatrix}&
M_{34} &= \begin{pmatrix}
0 & i & \frac{i}{2} & 0\\
0 & 0 & 0 & \frac{i}{2}\\
\frac{i}{2} & 0 & 0 & 0\\
0 & \frac{i}{2} & i & 0\end{pmatrix}\nn
M_{41} &= \begin{pmatrix}
-\frac{1}{2} & 0 & 0 & 1\\
0 & \frac{1}{2} & 0 & 0\\
0 & 0 & -\frac{1}{2} & 0\\
-1 & 0 & 0 & \frac{1}{2}\end{pmatrix}&
M_{42} &= \begin{pmatrix}
0 & -\frac{1}{2} & 1 & 0\\
\frac{1}{2} & 0 & 0 & 0\\
0 & 0 & 0 & -\frac{1}{2}\\
0 & -1 & \frac{1}{2} & 0\end{pmatrix}&
M_{43} &= \begin{pmatrix}
0 & \frac{i}{2} & i & 0\\
\frac{i}{2} & 0 & 0 & 0\\
0 & 0 & 0 & \frac{i}{2}\\
0 & i & \frac{i}{2} & 0\end{pmatrix}&
M_{44} &= \begin{pmatrix}
\frac{1}{2} & 0 & 0 & 1\\
0 & \frac{1}{2} & 0 & 0\\
0 & 0 & \frac{1}{2} & 0\\
1 & 0 & 0 & \frac{1}{2}\end{pmatrix}.
\end{align}
Now the overlap calculation $\langle \psi'| \psi\rangle$ reduces to taking the trace of a long product of matrices. Such tasks are performed in statistical mechanics by the transfer matrix method, and we can adopt the same strategy here to compute overlaps. 

% Starting with the simplest case of $\langle A | A \rangle$, we find
% %
% \ba\label{eq:OA}
% \ket{A} &=& (-1)^N \sum_{\{s\}} \textrm{Tr}[M_0^{(s_1)}\cdots M_0^{(s_N)}]\nn
% \braket{A|A}&=&\textrm{Tr}[M^N_{11}] =(3/2)^N .
% \ea

We begin by providing
a transfer matrix expression for the overlap between two bond coherent states, defined by Eq. (2) of the main text. %~(\ref{eq:bond-coherent-state}) 
From Eq.~(2) on expanding in terms of bond-product states (see Fig. 1) we have,
\ba
|\v N\rangle &=& \sum_{\{ \alpha_i\}} z_1^{\alpha_1} z_2^{\alpha_2}\cdots z_N^{\alpha_N} |N_1^{\alpha_1} N_2^{\alpha_2}\cdots N_N^{\alpha_N} \ra. 
\ea
Considering the overlap of $|\v N \ra$ with another bond coherent state 
$|\v M \ra= \sum_{\{ \beta_i\}} y_1^{\beta_1} y_2^{\beta_2}\cdots y_N^{\beta_N} |N_1^{\beta_1} N_2^{\beta_2}\cdots N_N^{\beta_N} \ra$ we have
\ba
\la \v M | \v N \ra = \sum_{\{ \alpha_i\},\{ \beta_i\} }  (z_1^{\alpha_1}\cdots z_N^{\alpha_ N})( \overline{y}_1^{\beta_1}\cdots \overline{y}_N^{\beta_N})
\la N_1^{\beta_1} N_2^{\beta_2}\cdots N_N^{\beta_N}|N_1^{\alpha_1} N_2^{\alpha_2}\cdots N_N^{\alpha_N}  \ra.
\ea
Equation (\ref{eq:mps-overlap}) allows us to express the overlap between two-bond product states as
\begin{equation}\label{eq:bondovl}
%\begin{split}
\la N_1^{\beta_1} N_2^{\beta_2}\cdots N_N^{\beta_N}|N_1^{\alpha_1} N_2^{\alpha_2}\cdots N_N^{\alpha_N}  \ra = \text{Tr}(\prod_i M_{\beta_i \alpha_i}). \\
%\end{split}
\end{equation}
%
% 
% The sixteen matrices which are  different from those in \cite{KPH19} due to different
% convention of singlet and triplet bond operators adopted in this paper can be easily derived from the ones presented in \cite{KPH19} and are not presented here for the sake of brevity. 
% Explicit expressions for the sixteen matrices $M_{\beta \alpha}, \beta,\alpha=1,..4$ can be found in Appendix A of
% \cite{KPH19}. Here we do not reproduce  them for the sake of brevity. It should be noted that the matrices used in \cite{KPH19} have to be modified by multiplicative factors of $1\sqrt{2}$ in order to
% conform to the convention used for defining singlet and triplet bond operators in this paper. 
Combining the last two equations we have
\ba
\la \v M | \v N \ra &=& \text{Tr}((\sum_{\alpha_1,\beta_1}{\overline{y}}_1^{\beta_1} z_1^{\alpha_1} 
M_{\beta_1 \alpha_1})\cdots (\sum_{\alpha_n,\beta_n}{\overline{y}}_n^{\beta_n} z_n^{\alpha_n} 
M_{\beta_n \alpha_n})) = \text{Tr}(\prod_i T_i)
\ea
with
\ba
\label{eq:transfer-mat}
T_i =\sum_{\alpha_i,\beta_i=1}^4{\overline{y}}_i^{\beta_i} z_i^{\alpha_i} 
M_{\beta_i \alpha_i}.
\ea

\subsection{Transfer matrix expression for energy density}
The bilinear-biquadratic (BLBQ) Hamiltonian is defined as
\ba 
H_i   = {\v S}_{i} \cdot {\v S}_{i+1}+\tan \tau~ ({\v S}_{i} \cdot {\v S}_{i+1})^2  = 1 + \tan \tau 
 -  (1 + 2\tan \tau)  {\cal{S}}_{i}^\dag {\cal{S}}_{i}+\tan \tau ({\cal{S}}_{i}^\dag {\cal{S}}_{i})^2 . 
\label{eq:Heisen-bond-operator-app}
\ea
In order to obtain a transfer matrix expression for the energy density we need to compute the action of the BLBQ Hamiltonian defined by 
Eq. (\ref{eq:Heisen-bond-operator-app}) in the main text on 
the bond coherent state $|\v N \ra$. This can be done by using the SB expression for the Hamiltonian given in Eq. (\ref{eq:Heisen-bond-operator-app}) by using standard SB techniques.
An important step in the computation is to utilize the fact that the operator ${\cal{S}}^{\dagger}_i{\cal{S}}_i$ affects only the bond operators at $(i-1)$th,  $i$ th and  $(i+1)$ th sites. For the purpose of this computation we introduce a new notation $| \bullet N^{\alpha}_{i-1} N^{\beta}_{i} N^{\gamma}_{i+1} \ra$ representing a state in which  for three consecutive sites $i-1$,  $i$  and $i+1$
the bond is one of the triplets or a  singlet  while for the other sites the bonds  are a mixture of the singlet and triplets. We have
%is identical to a bond-product state for the three sites $i-1$,$i$ and $i+1$ while for the other sites the bonds are a mixture of the singlet and triplet bonds. We have

\ba | \bullet N^{\alpha}_{i-1} N^{\beta}_{i} N^{\gamma}_{i+1} \ra = | \left(\prod_{j=1}^{i-2}N_j \right)  N^{\alpha}_{i-1} N^{\beta}_{i} N^{\gamma}_{i+1} \left(\prod_{j=i+2}^N N_j \right)\ra . \ea
Using the Schwinger boson expression for ${\cal{S}}_i$ defined in Eq. (1) of the main text we can derive
%~(\ref{eq:bond-coherent-state}) one can  derive,
%
\begin{equation}
%\begin{split}
 {\cal{S}}^{\dagger}_i{\cal{S}}_i \left( \sum_{\alpha, \beta, \gamma=1}^4 c_{\alpha \beta \gamma}|\v \bullet N^{\alpha}_{i-1} N^{\beta}_{i} N^{\gamma}_{i+1}\ra \right) =
 \sum_{\alpha,\beta} G^{\alpha, \beta}_i(\{c_{\alpha \beta \gamma} \})|\bullet N^{\alpha}_{i-1}S_i  N^{\beta}_{i+1} \ra, \\
\label{eq:coeffic-Hi-gen}
%\end{split}
\end{equation}
for  arbitrary complex coefficients  $c_{\alpha \beta \gamma}$ and functions $G^{\alpha, \beta}_i(\{c_{\alpha \beta \gamma} \})$. Lengthy  expressions of  $G^{\alpha, \beta}_i(\{c_{\alpha \beta \gamma} \})$ are delegated to the end of the supplementary material in the last section to maintain continuity of the discussion here.
Equation (\ref{eq:coeffic-Hi-gen}) allows us to  compute the action of the BLBQ Hamiltonian on the bond coherent state  as both $|\v N \ra$ and $ {\cal{S}}^{\dagger}_i{\cal{S}}_i |\v N \ra$ can be written in the form  $\sum_{\alpha, \beta, \gamma=1}^4 c_{\alpha \beta \gamma}|\v \bullet N^{\alpha}_{i-1} N^{\beta}_{i} N^{\gamma}_{i+1}\ra )$ using suitable coefficients $\{c_{\alpha \beta \gamma} \}$.  
The bond coherent state $|\v N \ra$ can be written as
\ba
|\v N \ra = \sum_{\alpha, \beta, \gamma=1}^4 z^{\alpha}_{i-1} z^{\beta}_{i} z^{\gamma}_{i+1} |\bullet N^{\alpha}_{i-1} N^{\beta}_{i} N^{\gamma}_{i+1} \ra .
\ea
Using  Eq. (\ref{eq:coeffic-Hi-gen})  with $c_{\alpha \beta \gamma} =z^{\alpha}_{i-1} z^{\beta}_{i} z^{\gamma}_{i+1}$ we have
\begin{equation}\label{eq:coeffic-Hi}
\begin{split}
{\cal{S}}^{\dagger}_i{\cal{S}}_i  |\v N \ra =& \sum_{\alpha,\beta} G^{\alpha, \beta}_i(\{ z^{\alpha}_{i-1} z^{\beta}_{i} z^{\gamma}_{i+1}  \})|\bullet N^{\alpha}_{i-1}S_i \v N^{\beta}_{i+1} \ra = \sum_{\alpha,\beta} J^{\alpha,\beta}_i |\bullet N^{\alpha}_{i-1}S_i  N^{\beta}_{i+1} \ra,
\end{split}
\end{equation}
with $J^{\alpha, \beta}_i$ defined as $G^{\alpha, \beta}_i(\{z^{\alpha}_{i-1} z^{\beta}_{i} z^{\gamma}_{i+1}  \})$. Again on applying ${\cal{S}}^{\dagger}_i{\cal{S}}_i$ on both sides of Eq. (\ref{eq:coeffic-Hi}) and using  Eq. (\ref{eq:coeffic-Hi-gen})  with $c_{\alpha \beta \gamma} =J^{\alpha,\gamma}_i \delta_{1,\beta}$ we have
\begin{equation}\label{eq:coeffic-Hi-1}
({\cal{S}}^{\dagger}_i{\cal{S}}_i)^2 | \v N \ra = \sum_{\alpha,\beta} G^{\alpha,\beta}_i(\{J^{\alpha,\gamma}_i \delta_{1,\beta}\}) |\bullet N^{\alpha}_{i-1}S_i  N^{\beta}_{i+1} \ra = \sum_{\alpha,\beta} K^{\alpha,\beta}_i |\bullet N^{\alpha}_{i-1}S_i  N^{\beta}_{i+1} \ra,
\end{equation}
with  $K^{\alpha, \beta}_i$ defined as $G^{\alpha,\beta}_i(\{J^{\alpha,\gamma}_i \delta_{1,\beta}\}) $. 

Using Eqs. (\ref{eq:coeffic-Hi}) and (\ref{eq:coeffic-Hi-1})  it is easy to  derive
the transfer matrix expression for the energy density corresponding to the Hamiltonian in 
Eq. (3) of the main text.  We have on using Eq. (\ref{eq:coeffic-Hi}) 

\ba
\frac{\la \v N  | H_i | \v N  \ra}{ \la \v  N  | \v N  \ra} = 1 + \tan \tau +   \sum_{\alpha,\beta} B^{\alpha,\beta}_i \la \v  N   | \bullet { N}^{\alpha}_{i-1} S_i {  N}^{\beta}_{i+1} \ra
\ea
with $B^{\alpha,\beta}_i = -  (1 + 2\tan \tau)J^{\alpha, \beta}_i +  \tan \tau K^{\alpha, \beta}_i $.
The computation of  $\la \v N  | \bullet N^{\alpha}_{i-1} S_i \v N^{\beta}_{i+1} \ra$ gets simplified by  the  recognition that the state $|\bullet N^{\alpha}_{i-1}S_i  N^{\beta}_{i+1} \ra$ is
a special case of a bond-coherent state defined by Eq.~(2) of the main text with mixing coefficients corresponding to the $(i-1)$th, $i$th
and $(i+1)$th bonds given by $z_{i-1}^{\gamma_{i-1}}= \delta_{\gamma_{i-1},\alpha}$ , $z_{i}^{\gamma_i}= \delta_{\gamma_i,1}$ and  $z_{i+1}^{\gamma_{i+1}}= \delta_{\gamma_{i+1},\beta}$. Hence, on using Eq. (\ref{eq:bondovl}) we have
\ba 
&& \sum_{\alpha,\beta} B^{\alpha, \beta}_i \la \v  N  | \bullet {N}^{\alpha}_{i-1} S_i {  N}^{\beta}_{i+1} \ra \nn
&& =\sum_{\alpha,\beta} B^{\alpha, \beta}_i  \text{Tr} \Bigl(\prod_{j= i+2,N} {X}_j\prod_{j= 1,i-2} {X}_j  (\sum_{\beta_{i-1}} \overline{z}_{i-1}^{\beta_{i-1}} M_{\beta_{i-1} \alpha}) (\sum_{\beta_{i}} \overline{z}_{i}^{\beta_{i}} M_{\beta_{i} 1}) (\sum_{\beta_{i+1}} \overline{z}_{i+1}^{\beta_{i+1}} M_{\beta_{i+1} \beta})\Bigl),
\ea
%with {X}_j = \sum_{\alpha_j,\beta_j=1}^4 M_{\beta_j \alpha_j} {\overline{z}}_j^{\beta_j} z_j^{\alpha_j}.
 with
 \ba
{X}_j &=& \sum_{\alpha_j,\beta_j=1}^4 M_{\beta_j \alpha_j} {\overline{z}}_j^{\beta_j} z_j^{\alpha_j}. \label{eq:tmat}
\ea
Noting the fact that $B^{\alpha, \beta}_i$ is solely a function of $z^\alpha_{i-1}$, $z^\alpha_{i}$, $z^\alpha_{i+1}$ for $\alpha=1,..4$ we can combine the above equation into the following transfer matrix expression
\ba
\sum_{\alpha,\beta} B^{\alpha, \beta}_i \la \v  N  |\bullet  { N}^{\alpha}_{i-1} S_i {  N}^{\beta}_{i+1} \ra
= \text{Tr} ((\prod_{j=1}^{i-2} {X}_j){P}_i (\prod_{j=i+2}^{N} {X}_j)), 
\ea
with
\ba
{P}_i = \sum_{\alpha,\beta, \beta_{i-1}, \beta_{i},\beta_{i+1}} \overline{z}_{i-1}^{\beta_{i-1}} \; \overline{z}_{i}^{\beta_{i}} \; \overline{z}_{i+1}^{\beta_{i+1}}
B^{\alpha, \beta}_i M_{\beta_{i-1} \alpha} \; M_{\beta_{i} 1} \;  M_{\beta_{i+1} \beta}.
\label{eq:Amat}
\ea
The energy density corresponding to the Hamiltonian defined by Eq. \ref{eq:Heisen-bond-operator-app} in the main text  is thus
%\ref{eq:Heisen-bond-operator-app}
\ba
\label{Eq:eden}
\frac{\la \v N  | H_i | \v N  \ra}{ \la \v  N  | \v N  \ra} =1 +\tan \tau + \frac{  \text{Tr} ((\prod_{j=1}^{i-2} {X}_j){P}_i (\prod_{j=i+2}^{N} {X}_j))}{ \text{Tr}(\prod_{j} {X}_j)}.
\ea
% with, 
% \begin{equation}\label{eq:Kmat}
% \begin{split}
% {\bf{D}}_i &= \\ & \sum_{\alpha,\beta, \beta_{i-1}, \beta_{i},\beta_{i+1}} \overline {z}_{i-1}^{\beta_{i-1}} \; \overline{z}_{i}^{\beta_{i}} \; \overline{z}_{i+1}^{\beta_{i+1}}
% K^{\alpha, \beta}_i M_{\beta_{i-1} \alpha} \; M_{\beta_{i} 1} \;  M_{\beta_{i+1} \beta}.
% \end{split}
% \end{equation}
\subsection{Saddle point analysis}
As we mentioned in the main text we consider two kinds of ansatz, the uniform ansatz and the staggered ansatz. For the uniform ansatz we have for all bonds, $N_i=z^1 N^1 + z^4N^4 $ while for the staggered ansatz 
we have $N_i=z^1 N^1 + (-1)^i z^4N^4 $ with $z^1=\cos \frac{\theta}{2}$ and $z^4=e^{i \phi} \sin \frac{\theta}{2}$. For the staggered ansatz   denoting the transfer matrix defined by Eq.~(\ref{eq:tmat}) for even sites as $X_e$ and for odd sites as $X_o$ we have
\begin{equation}
X_e= \begin{pmatrix}
\frac{1}{2}(1- \cos \phi \sin \theta) & 0 & 0 & (1+ \cos \phi \sin \theta) \\
0 & \frac{1}{2}(-\cos \theta - i \sin \phi \sin \theta) & 0 & 0 \\
0 & 0 & \frac{1}{2}(-\cos \theta + i \sin \phi \sin \theta) & 0 \\
1-\cos \phi \sin \theta & 0 & 0 & \frac{1}{2}(1+\cos \phi \sin \theta )
\end{pmatrix}
\end{equation}
and $X_o = X_e(-\theta,\phi)$.  We also denote the $P$ matrix defined by Eq. (\ref{eq:Amat}) for the staggered case as $P_s$ and the uniform case as $P_u$. Then we have, for even sites,
\ba
\frac{\la \v N  | H_i | \v N  \ra}{ \la \v  N  | \v N  \ra} &=&1 +\tan \tau +
\frac{ \text{Tr}(P_s {(X_eX_o)}^{\frac{N-4}{2}}X_e) }{ \text{Tr}( {(X_eX_o)}^{\frac{N}{2}} )}~~~~~~~~(\mbox{staggered}) \nn
&=& 1 +\tan \tau +
\frac{ \text{Tr}(P_u {X_e}^{N-3}) }{ \text{Tr}(X_e^{N} )}~~~~~~~~~~~~~~~~~~(\mbox{uniform}) 
\ea
while for odd sites for the staggered case denoting the $P$ matrix by $P'_s$ we have
\ba
\frac{\la \v N  | H_i | \v N  \ra}{ \la \v  N  | \v N  \ra} &=&1 +\tan \tau +
\frac{ \text{Tr}(X_o P'_s {(X_oX_e)}^{\frac{N-4}{2}}) }{ \text{Tr}( {(X_eX_o)}^{\frac{N}{2}} )}.
\ea
We have checked that the energies are the same for the BLBQ Hamiltonian at even and odd sites for the staggered ansatz even though the transfer matrix expressions for the energies look somewhat different. The reason for the equality is as follows. The BLBQ Hamiltonian is invariant under identical spin rotation on all sites which generates a rotation in the triplet sector while leaving the singlet sector invariant. Hence, as a staggered ansatz $N_i=z^1 N^1 + (-1)^i z^4N^4 $ gets transformed to  $N_i=z^1 N^1 + (-1)^{(i+1)} z^4N^4 $ under such a rotation which takes $N^4 \rightarrow -N^4$ leaving $N^j$ for $j \ne 4$ invariant, the energies for the staggered ansatz for Hamiltonian $H_i$ for an even or odd sites $i$ are exactly equal. The above equations are used for  drawing conclusions regarding the nature of the saddle point for the uniform and staggered ansatzes, as mentioned in the main text. Fig 1(a) in the main text is also produced using these equations. 

\subsection{$\braket{S_i^z}$ calculation}
In this section we describe how to calculate $\braket{S_i^z}$ for $\ket{\v N}$ and include the graph of it. Transfer matrix method outlined in the above is applied for $\braket{S_i^z}$ again except $H_i$ is changed to $S_i^z$. Since $S_i^z$ is only affecting $N_{i-1}$ and  $N_i$ bond, we will calculate $S_i^z N_{i-1}N_{i-1}\ket{v}$ and the result is
\ba
S_i^z N_{i-1}N_{i-1}\ket{v}&=&\Bigl(-\frac{1}{2}N^4_{i-1}N^1_{i}+\frac{1}{2}N^1_{i-1}N^4_{i}+(-1)^i z^1 z^4 N^1_{i-1} N^1_{i}+(-1)^{i+1} z^1 z^4 N^4_{i-1} N^4_{i}\Bigl)\ket{v}.
\ea
Then $\braket{\v N|S_i^z|\v N}$ is organized as
\ba
\braket{\v N|S_i^z|\v N}&=&-\frac{1}{2}\braket{\v N|\bullet N_{i-1}^4N_{i}^1}+\frac{1}{2}\braket{\v N|\bullet N_{i-1}^1 N_{i}^4}+(-1)^i z^1 z^4\braket{\v N|\bullet N_{i-1}^1 N_{i}^1}+(-1)^{i+1}z^1z^4\braket{\v N|\bullet N_{i-1}^4N_{i}^4}\nn
\ea
where $| \bullet N^{\alpha}_{i-1} N^{\beta}_{i} \ra$ is defined as
\ba
| \bullet N^{\alpha}_{i-1} N^{\beta}_{i}\ra = | \left(\prod_{j=1}^{i-2}N_j \right)  N^{\alpha}_{i-1} N^{\beta}_{i} \left(\prod_{j=i+1}^N N_j \right)\ra . \ea
Now $\braket{\v N|S_i^z|\v N}$ and $\braket{\v N|\v N}$ are analytically solvable using transfer matrix method described previously. We find
\ba
\braket{S_i^z}&=&(-1)^i\frac{2 \sqrt{2} \sin \theta}{\sqrt{5-3 \cos 2 \theta}}
\ea
and we can easily verify $\braket{S_{i}^z}=(-1)^i$ for $\theta=\frac{\pi}{2}$, where $N_i$ becomes
\ba
N_i&=&a_i^{\dagger}b^{\dagger}_{i+1}~~~(\text{even}~i)\nn
&=&b_i^{\dagger}a^{\dagger}_{i+1}~~~(\text{odd}~i),
\ea
and there are two $a^{\dagger}$ for even sites and two $b^{\dagger}$ for odd sites.
\begin{figure*}[t]
\centering
\includegraphics[width=50mm]{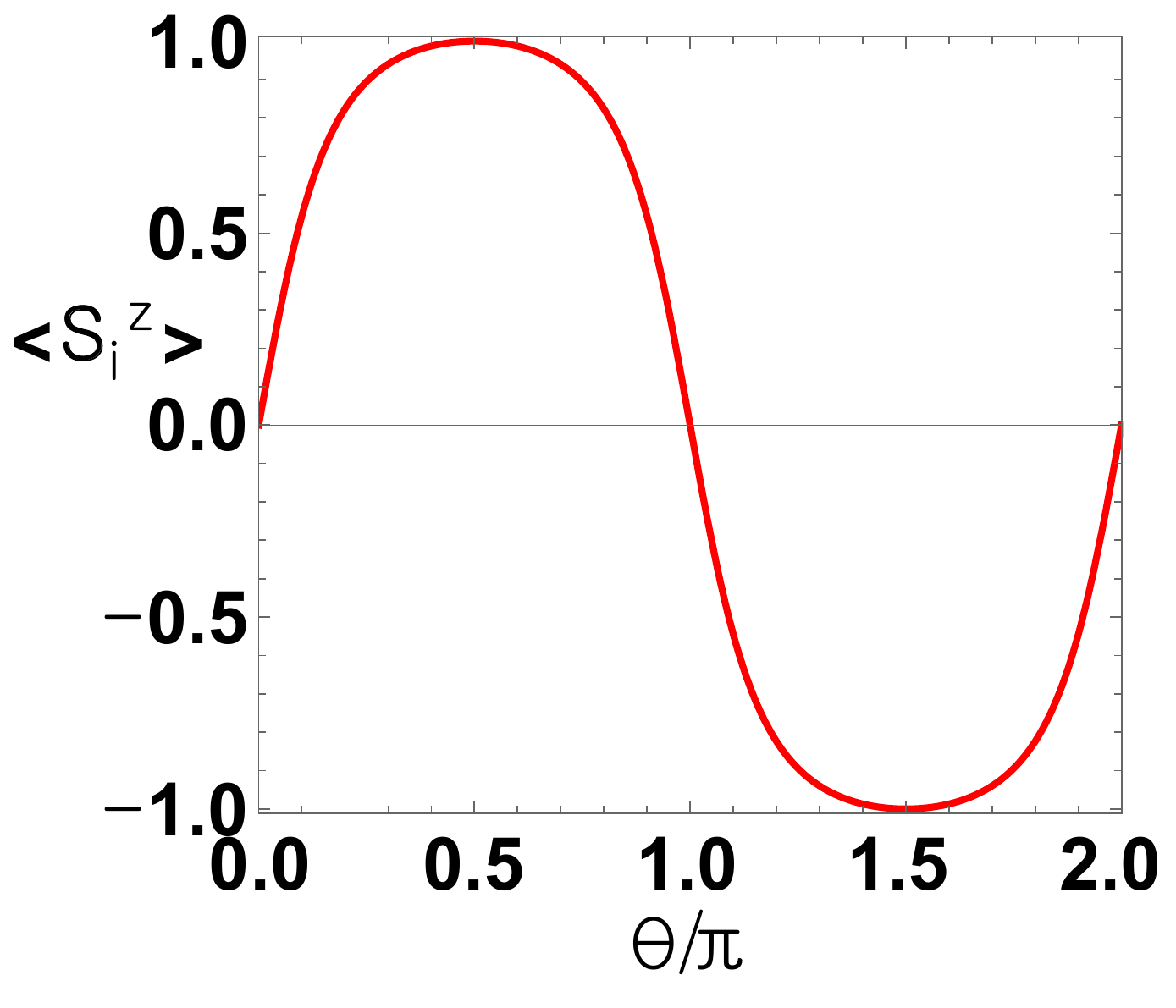}
\caption{$\braket{S_i^z}$ graph for even $i$} \label{fig:Sz graph}
\end{figure*}

\section{Energy functional}
In this section we provide details of the calculation of the long-range energy functional described in Eq. (12) of the main text obtained by performing saddle point expansion about the AKLT point. As explained in the main text we assume that 
the fluctuations about saddle point primarily come from triplet bonds replacing  a small number singlet bonds in the AKLT state. Hence, we consider $z^{1}_i=1$ for all sites and expand the numerator and denominator of the energy density  keeping  terms up to second order
in $z^{\alpha}_i$, for  $i=1,..N$ and $\alpha \ne 1$. The transfer matrix corresponding to the AKLT state is given by the matrix $M_{11}$ [see also Eq.~(\ref{eq:transfer-mat})]. 

From the approach of Eq.(9) in the main text, if we keep until the second order of $z$'s, $\braket{\v N|\v N}$ is expressed as 
\ba\label{eq:approxexp}
\braket{\v N|\v N}&=&\braket{A|A}+\sum_i \Bigl(\braket{A|\mathcal{T}_i}+\braket{\mathcal{T}_i|A}\Bigl) +\sum_{j<i}\Bigl(\braket{A|\mathcal{T}_j\mathcal{T}_i}+\braket{\mathcal{T}_j\mathcal{T}_i|A}\Bigl)+\sum_{i,j}\braket{\mathcal{T}_i|\mathcal{T}_j},
\ea
and each term in Eq. (\ref{eq:approxexp}) calculated by transfer matrix method gives
\ba
& & \braket{A|A}=\Bigl(\frac{3}{2}\Bigl)^N , ~~~~~~~~~~~~~ \sum_i\Bigl(\braket{A|\mathcal{T}_i}+\braket{\mathcal{T}_i|A}\Bigl)=0\nn 
&& \sum_{j<i}\Bigl(\braket{A|\mathcal{T}_j\mathcal{T}_i}+\braket{\mathcal{T}_j\mathcal{T}_i|A}\Bigl)=\Bigl(\frac{3}{2}\Bigl)^N\frac{1}{2}\sum_{i\neq j}f_{ij}(z_i^\alpha z_j^\alpha+\overline{z}_i^\alpha \overline{z}_j^\alpha) \nn
&& \sum_{i,j}\braket{\mathcal{T}_i|\mathcal{T}_j}=\Bigl(\frac{3}{2}\Bigl)^N\Bigl[\frac{1}{2}\sum_{i\neq j}f_{ij}(z_i^\alpha \overline{z}_j^\alpha+\overline{z}_i^\alpha z_j^\alpha)+\sum_{i,\alpha}|z_i^\alpha|^2\Bigl].
\ea
After organization the above results, $\braket{\v N|\v N}$ becomes
\ba\label{eq:approx result}
\braket{\v N|\v N}&=&\Bigl(\frac{3}{2}\Bigl)^N\Bigl[1+\sum_{i,\alpha}|z_i^\alpha|^2+2\sum_{i\neq j,\alpha}f_{ij}x_i^\alpha x_j^\alpha\Bigl].
\ea
Similarly, we can evaluate
\ba
\braket{\partial_t \v N|\v N}&=&\sum_i \Bigl(\braket{\partial_t\mathcal{T}_i|A}\Bigl) +\sum_{i\neq j} \braket{(\partial_t\mathcal{T}_i)\mathcal{T}_j|A}+\sum_{i,j}\braket{\partial_t \mathcal{T}_i|\mathcal{T}_j} =\Bigl(\frac{3}{2}\Bigl)^N\sum_{i,j,\alpha}f_{ij}(\partial_t\overline{z}_i^\alpha)z_j^\alpha\nn
\braket{\v N|H_i|\v N}&=&\braket{A|H_i|A}+\sum_j \Bigl(\braket{A|H_i|\mathcal{T}_j}+\braket{\mathcal{T}_j|H_i|A}\Bigl)+\sum_{j<k}\Bigl(\braket{A|H_i|\mathcal{T}_j\mathcal{T}_k}+\braket{\mathcal{T}_j\mathcal{T}_k|H_i|A}\Bigl)+\sum_{j,k}\braket{\mathcal{T}_j|H_i|\mathcal{T}_k}\nn
&=&\Bigl(\frac{3}{2}\Bigl)^N\Bigl[-\frac{7}{3}+\tan \tau+\sum_{j<k<i,i<j<k}\Bigl((-5+9\tan \tau)\eta^{j-k+N}+(-\frac{7}{3}+\tan \tau)\eta^{k-j}\Bigl)4 x_j^\alpha x_k^\alpha\nn
&+&\sum_{j<i<k}\Bigl((-5+9\tan \tau)\eta^{k-j}+(-\frac{7}{3}+\tan \tau)\eta^{-j+k-N}\Bigl)4 x_j^\alpha x_k^\alpha\nn
&+&\sum_{k\neq j}(-\frac{7}{3}+\tan \tau)f_{kj}4 x_j^\alpha x_k^\alpha+\sum_{j\neq i}(-\frac{7}{3}+\tan \tau)|z_j^\alpha|^2+(-\frac{5}{9}+\frac{1}{9}\tan \tau)|z_i^\alpha|^2\Bigl] . 
\ea

To calculate $-i\braket{\partial_t \v N|\v N}/\braket{\v N|\v N}$ and $\braket{\v N|H_i|\v N}/\braket{\v N|\v N}$, we note
\ba
\braket{\v N|\v N}^{-1} \approx \Bigl(\frac{2}{3}\Bigl)^N(1-\sum_{i,\alpha}|z_i^\alpha|^2-2\sum_{i\neq j,\alpha}f_{ij}x_i^\alpha x_j^\alpha) .
\ea
Equations (11) and (12) in the main text are recovered after keeping terms up to  second order in $z$'s.

\section{Explicit expressions for $G_i^{\alpha, \beta}$}
\label{append:Gab}
Here, we provide  explicit expressions for $G^{\alpha, \beta}_i(\{c_{\alpha \beta \gamma} \})$  defined in section III.

% $J_i^{\alpha, \beta}$ defined in section II. $G^{\alpha, \beta}_i(\{c_{\alpha \beta \gamma} \})$ can be obtained  by replacing $z^{\alpha}_{i-1} z^{\beta}_{i} z^{\gamma}_{i+1}$ with $c_{\alpha \beta \gamma} $ in the following equations. We have 
\ba
G_i^{11} &=& {9 \over 4}  c_{111} - 
 {3 \over 4} \Bigl( c_{221}  +
  c_{122} + c_{331}  + c_{133} +  c_{441}  + c_{144}\Bigr) 
 + 
 {i \over 4 }  \Bigl(  c_{342}  -  c_{243}  -  c_{432}  +  c_{423}  +  c_{234}  -  c_{324} \Bigr)    
 + {1 \over 4} \Bigl(
 c_{212}  +   c_{313}  +   c_{414} \Bigr) , \nonumber
\ea
\ba
G_i^{12} &=& - 
 {9 \over 4 \sqrt{2}}  c_{112} + 
 {9 i \over 4 \sqrt{2}} c_{114}
+ {3 \over 4 \sqrt{2}}  \Bigl(  c_{121}    +  c_{222} +   c_{332} - 
 c_{132}    +  c_{123} +   c_{442} \Bigr) 
 + {3 i  \over 4 \sqrt{2}} \Bigl(
  c_{134}  
 -   c_{141}   -   c_{143}  - c_{224}  - c_{334}  -c_{444} \Bigr) \nn
   &+& {1 \over 4 \sqrt{2}} \Bigl(c_{321}  - c_{231}  + c_{312}  - c_{213}  - c_{233} - c_{211} +  c_{323}- c_{244} +  c_{424} \Bigr) 
 + {i \over 4 \sqrt{2} } \Bigl(
 c_{411}
 - c_{242}  - c_{341}   -   c_{343}   + c_{422} +  c_{431}\nn &&~~~     +  
  c_{413}   +  c_{433}   - c_{314} \Bigr) , \nonumber
\ea
\ba
G_i^{13} &=& {9 \over 4} c_{113}-{3 \over 4} \Bigl(c_{131}  +  c_{223}+ c_{333} +c_{443}  \Bigr) + 
 {3i \over 4}   \Bigl( c_{124}
-  c_{142}  \Bigr) 
+ {1 \over 4} \Bigl(
c_{311}     - 
 c_{232}  +  c_{322}  +  
 c_{344}-  
c_{434}\Bigr) \nn
 &+& {i \over 4} \Bigl(
 c_{421} -   c_{241} + c_{412} -  c_{214}\Bigr) , \nonumber
\ea%\sqrt{2}i
\ba
 G_i^{14} &=& 
 {9 \over 4 \sqrt{2}}  c_{112} + 
 {9 i \over 4 \sqrt{2}} c_{114}  
 + {3 \over 4 \sqrt{2}} \Bigl(  c_{123} -  c_{121}  
- c_{222} -  c_{132}-   c_{332}  - c_{442}\Bigr) 
  + {3 i \over 4 \sqrt{2} } \Bigl(
  c_{143}-   c_{141}-   c_{224} - c_{134}  - c_{334} - c_{444} + 
 c_{211}\Bigr) \nn
 & +& {1 \over 4 \sqrt{2}} \Bigl(
  c_{321} -   c_{231} +   c_{312}    - c_{213}  +  c_{233} -  c_{323}  +  c_{244} -  c_{424}\Bigr) 
 + {i \over 4 \sqrt{2} } \Bigl(
c_{341} -    c_{242} -  c_{343} +  c_{411}+ c_{422}  -  c_{431}-   c_{413} \nn
  &&~~~ +   c_{433} +  c_{314}\Bigr) ,   \nonumber
\ea
\ba
G_i^{21} &=& 
 - {9 \over 4 \sqrt{2}} c_{221}  + 
{ 9 i \over 4 \sqrt{2}} c_{411}  
 + {3 \over 4 \sqrt{2}} \Bigl( c_{121} + c_{222}+ c_{231} - c_{321} +  c_{233}+ c_{441}\Bigr) 
 + {3 i \over 4 \sqrt{2}} \Bigl(
  c_{341} - c_{141} - c_{422} - c_{431}  -c_{433} -  c_{444}\Bigr) \nn
&+& {1 \over 4 \sqrt{2} } \Bigl(
  c_{132} -
 c_{112} + c_{312} - c_{332}-c_{123}-c_{213}+ c_{323} -  
   c_{442} +
 c_{424}\Bigr)  + {i \over 4 \sqrt{2}} \Bigl(c_{143} - c_{242} - c_{343} + c_{413} + c_{114}+  c_{224}-  c_{134} \nn
  &&~~~ -  c_{314} + c_{334}\Bigr) , \nonumber
\ea  %
\ba
G_i^{22} &=& 
  c_{212} - 
  c_{414}- 
 i c_{412} - 
 i c_{214} 
+ {1 \over 2} \Bigl(
c_{322}  
  - c_{221}  -  c_{122} - c_{223}+ c_{441} + c_{443} + c_{144} - c_{344} \Bigr)
+
 {i \over 2 }\Bigl(c_{241} +  c_{142}- c_{342}+  c_{243}+ c_{421} \nn 
&&~~~   + c_{423}  + c_{124}- c_{324}\Bigr) , \nonumber
\ea

\ba
 G_i^{23} &=& - 
 {9 \over 4 \sqrt{2} }  c_{213}+ 
 {9 i \over 4 \sqrt{2} }  c_{413} 
 + {3 \over 4 \sqrt{2}}  \Bigl(
    c_{231} +  c_{123} + c_{233}- c_{323}- c_{424}+  c_{442}  \Bigr) 
  + {3 i \over 4 \sqrt{2}} \Bigl(
  c_{242}  - c_{143}  +  c_{343} - c_{431} - c_{433} - c_{224} \Bigr) \nn 
&+& {1 \over 4 \sqrt{2}} \Bigl(
 c_{132}   - c_{121}  - 
 c_{211}   - 
 c_{112}  -  
 c_{222}   +  
  c_{321} +  
c_{312}   -  
 c_{332}  -  c_{244}\Bigr)  
 + {i \over 4 \sqrt{2}} \Bigl(
 c_{141}    -  c_{341}   + c_{411}  +  
  c_{422}   +   c_{114}  -  c_{134} -  c_{314} \nn
&&~~~  +   c_{334}  + c_{444} \Bigr) , \nonumber
\ea
\ba
 G_i^{24} &=&  - {5 \over 4} \Bigl(
 c_{414}+ c_{212} \Bigr)  +
 {5i \over 4 } \Bigl(c_{412}  - c_{214} \Bigr)
 + {3 \over 4} \Bigl(
 c_{232} +  c_{434} \Bigr) + 
 {3i \over 4 } \Bigl(c_{234} - c_{432}  \Bigr) 
 + {1 \over 4} \Bigl(
 c_{122}- c_{111} +
  c_{221} +
 c_{131} +
 c_{311}  -
c_{322} \nn &&~~~ -
 c_{331} + c_{113}  -
 c_{223} -c_{133} - c_{313} +c_{333} +c_{441}  -c_{443}  +c_{144} -c_{344}  \Bigr) 
 + {i \over 4} \Bigl( c_{241} - c_{142}+ c_{342}  - c_{243} \nn  &&~~~  - c_{421}  + c_{423}  +c_{124} - c_{324}  \Bigr) , \nonumber
\ea
\ba
G_i^{31} &=& {9 \over 4} 
c_{311} -{3 \over 4} \Bigl( c_{131}   + c_{333} +c_{322} +c_{344}\Bigr) 
 + {3i \over 4} \Bigl( c_{241}  - 
 c_{421} \Bigr) 
 +{1 \over 4} \Bigl(
c_{113} -  
  c_{232}   + 
c_{223}   +  
 c_{443} -  
 c_{434} \Bigr)  
 + {i \over 4}\Bigl( 
 c_{142}  + 
c_{412} \nn  &&~~~- 
c_{124} - 
c_{214}  \Bigr) ,  \nn
 G_i^{32} &=&
   - {9 \over 4 \sqrt{2}} c_{312}  +{9 i  \over 4 \sqrt{2}}  c_{314}  
  + {3 \over 4 \sqrt{2} } \Bigl(
 c_{132}  + 
  c_{321}   - c_{332}   + c_{323} - c_{244} + c_{424} \Bigr)\nn
 &+& {3 i \over 4 \sqrt{2}} \Bigl(c_{422}  -  c_{242} -c_{341} -  c_{343}-  c_{134} +   c_{334}\Bigr) 
 +
{1 \over 4 \sqrt{2} }\Bigl(c_{121}  +
c_{211}  +
c_{112} +
c_{222} +
 c_{231}  +
 c_{123} +
 c_{213} +
 c_{233} +
 c_{442} \Bigr)\nn &-& {i \over 4 \sqrt{2}} \Bigl(
  c_{141}  +
  c_{143} +
  c_{411}  +
  c_{431}  +
  c_{413}  +
  c_{433}  +
  c_{114} +
  c_{224} \nn
  &&~~~ +
 c_{444} \Bigr) , \nn
     G_i^{33} &=&
    {9 \over 4} 
 c_{313} - {3 \over 4} \Bigl(
 c_{331}  +
 c_{133} \Bigr) + 
 {3 i \over 4 } \Bigl( c_{243} -  c_{342}  - c_{423}  +c_{324} \Bigr) 
 +{1 \over 4} \Bigl( c_{111} +
 c_{221}  +
 c_{122} +
 c_{212} +
 c_{441}   +
 c_{144} +
 c_{414}\Bigr) \nn
 &+&
 {i \over 4} \Bigl(c_{234} -  c_{432} \Bigr) , \nonumber
\ea
\ba
%i\sqrt{2}
   G_i^{34} &=&
 {9 \over 4 \sqrt{2}}c_{312}  + 
 {9 i \over 4 \sqrt{2}} c_{314} 
 + {3 \over 4 \sqrt{2} } \Bigl(
 c_{323} - c_{132} -c_{321} - c_{332}   -c_{244} +c_{424} \Bigr)  +  {3 i \over 4 \sqrt{2}} \Bigl(
 c_{242} - c_{341}  +c_{343}  - c_{422}  -  c_{134} -  c_{334} \Bigr) \nn
&+& {1 \over 4 \sqrt{2} } \Bigl(c_{121}  +
c_{211}  +
c_{112} +
 c_{222} -
 c_{231}  -
 c_{123} -
 c_{213} 
 +
 c_{233} 
  + c_{442} \Bigr)   + {i \over 4 \sqrt{2} } \Bigl( 
  c_{141} -
 c_{143}  +
 c_{411}  -
 c_{431}-
 c_{413} \nn  &+&
  c_{433} +
 c_{114} 
 +
 c_{224} 
 + c_{444} \Bigr) , \nonumber
\ea
\ba
 G_i^{41} &=&
      {9 \over 4 \sqrt{2}}c_{211}  +
{ 9 i \over 4 \sqrt{2}} c_{411} 
+ {3 \over 4 \sqrt{2}} \Bigl(
 c_{213}
     -  c_{121}  - c_{222} - c_{321}  - c_{233} - c_{244} \Bigr) 
 + {3 i \over 4 \sqrt{2} } \Bigl(
 c_{431} - c_{141} -  c_{341} -c_{422} -  c_{433}  -c_{444} \Bigr) \nn
&+& {1 \over 4 \sqrt{2} } \Bigl(
 c_{112} +
 c_{132} +
c_{312} +
 c_{332} -
 c_{123} -
 c_{213} -
 c_{323} +
c_{442}    -
 c_{424} \Bigr)  + {i \over 4 \sqrt{2}} \Bigl(
 c_{114} - c_{242}   -
  c_{143}   -
  c_{343}   -
  c_{413}   +
 c_{224}  +  
 c_{134} \nn  +  &&~~~c_{314} +  c_{334} \Bigr) \nonumber, 
\ea 
\ba
G_i^{42} &=&   {5 \over 4} \Bigl(
  c_{414}+ c_{212} \Bigr)  -
 {5i \over 4 } \Bigl( c_{412}  - c_{214} \Bigr) 
 + {3 \over 4} \Bigl(
 c_{232} + c_{434} \Bigr) + 
 {3i \over 4 } \Bigl(c_{234} -  c_{432}  \Bigr) 
 + {1 \over 4} \Bigl(
 - c_{221} +c_{111} -
 c_{122} +
 c_{131} +
 c_{113}  -
c_{223}  \nn &&~~~ +
 c_{133} +
 c_{311} 
  - c_{223} 
+
 c_{331} + 
c_{313} +
 c_{333} +
 c_{144}  -
c_{443} -
c_{441} -
 c_{443}  \Bigr) + {i \over 4} \Bigl(-c_{142} + c_{241}+c_{243}  -c_{342}  \nn
&&~~~ +  c_{124}  +c_{324}  - c_{421} - c_{423} \Bigr) , \nonumber
\ea
\ba
  G_i^{43} &=&
-  {9 \over 4 \sqrt{2}}c_{213}  - 
 {9 i \over 4 \sqrt{2}}c_{413} 
 + {3 \over 4 \sqrt{2} } \Bigl(
  c_{323} + c_{231} + c_{123} -  c_{233}  - c_{442} + c_{424} \Bigr) 
 +  {3 i \over 4 \sqrt{2}} \Bigl(
 c_{242} +c_{143}  + c_{343}  - c_{224}  +  c_{431} -  c_{433} \Bigr) \nn
&+& {1 \over 4 \sqrt{2} } \Bigl( c_{121}  +
 c_{112}  +
 c_{211} +
  c_{222}+
 c_{132}  +
 c_{321} +
 c_{312} 
 +
 c_{332} 
+ c_{244} \Bigr) 
 + {i \over 4 \sqrt{2} } \Bigl( 
  c_{141} -
  c_{341}  +
 c_{114} +
 c_{134}-
 c_{314}  +
  c_{334} \nn &&~~~ +
 c_{411} 
 +
  c_{422} 
  +
  c_{444} \Bigr) , \nonumber
\ea
\ba
G_i^{44} &=&
c_{212}  - 
c_{414}+ 
 i c_{412}  + 
 i c_{214} 
 + {1 \over 2}\Bigl(
  c_{441} - 
 c_{122} -
 c_{322}  +
c_{223}
  - c_{221}  -
 c_{443}  +  
c_{144}  +  
 c_{344} \Bigr) 
 +
 {i \over 2 } \Bigl(c_{243}
 - 
c_{241}  - 
 c_{142} \nn &&~~~ - 
 c_{342}  - 
  c_{421}  + 
c_{423}    - 
 c_{124} - 
 c_{324}  \Bigr) . \nonumber
\ea
\end{widetext}

\bibliography{fmps}

\begin{thebibliography}{17}%
\makeatletter
\providecommand \@ifxundefined [1]{%
 \@ifx{#1\undefined}
}%
\providecommand \@ifnum [1]{%
 \ifnum #1\expandafter \@firstoftwo
 \else \expandafter \@secondoftwo
 \fi
}%
\providecommand \@ifx [1]{%
 \ifx #1\expandafter \@firstoftwo
 \else \expandafter \@secondoftwo
 \fi
}%
\providecommand \natexlab [1]{#1}%
\providecommand \enquote  [1]{``#1''}%
\providecommand \bibnamefont  [1]{#1}%
\providecommand \bibfnamefont [1]{#1}%
\providecommand \citenamefont [1]{#1}%
\providecommand \href@noop [0]{\@secondoftwo}%
\providecommand \href [0]{\begingroup \@sanitize@url \@href}%
\providecommand \@href[1]{\@@startlink{#1}\@@href}%
\providecommand \@@href[1]{\endgroup#1\@@endlink}%
\providecommand \@sanitize@url [0]{\catcode `\\12\catcode `\$12\catcode
  `\&12\catcode `\#12\catcode `\^12\catcode `\_12\catcode `\%12\relax}%
\providecommand \@@startlink[1]{}%
\providecommand \@@endlink[0]{}%
\providecommand \url  [0]{\begingroup\@sanitize@url \@url }%
\providecommand \@url [1]{\endgroup\@href {#1}{\urlprefix }}%
\providecommand \urlprefix  [0]{URL }%
\providecommand \Eprint [0]{\href }%
\providecommand \doibase [0]{http://dx.doi.org/}%
\providecommand \selectlanguage [0]{\@gobble}%
\providecommand \bibinfo  [0]{\@secondoftwo}%
\providecommand \bibfield  [0]{\@secondoftwo}%
\providecommand \translation [1]{[#1]}%
\providecommand \BibitemOpen [0]{}%
\providecommand \bibitemStop [0]{}%
\providecommand \bibitemNoStop [0]{.\EOS\space}%
\providecommand \EOS [0]{\spacefactor3000\relax}%
\providecommand \BibitemShut  [1]{\csname bibitem#1\endcsname}%
\let\auto@bib@innerbib\@empty
%</preamble>
\bibitem [{\citenamefont {Auerbach}(1994)}]{Auerbach}%
  \BibitemOpen
  \bibfield  {author} {\bibinfo {author} {\bibfnamefont {A.}~\bibnamefont
  {Auerbach}},\ }\href@noop {} {\emph {\bibinfo {title} {Interacting Electrons
  and Quantum Magnetism}}}\ (\bibinfo  {publisher} {Springer-Verlag},\ \bibinfo
  {year} {1994})\BibitemShut {NoStop}%
\bibitem [{\citenamefont {Haldane}(1983)}]{Haldane83}%
  \BibitemOpen
  \bibfield  {author} {\bibinfo {author} {\bibfnamefont {F.~D.~M.}\
  \bibnamefont {Haldane}},\ }\href {\doibase 10.1103/PhysRevLett.50.1153}
  {\bibfield  {journal} {\bibinfo  {journal} {Phys. Rev. Lett.}\ }\textbf
  {\bibinfo {volume} {50}},\ \bibinfo {pages} {1153} (\bibinfo {year}
  {1983})}\BibitemShut {NoStop}%
\bibitem [{\citenamefont {Affleck}\ \emph {et~al.}(1987)\citenamefont
  {Affleck}, \citenamefont {Kennedy}, \citenamefont {Lieb},\ and\ \citenamefont
  {Tasaki}}]{AKLT87}%
  \BibitemOpen
  \bibfield  {author} {\bibinfo {author} {\bibfnamefont {I.}~\bibnamefont
  {Affleck}}, \bibinfo {author} {\bibfnamefont {T.}~\bibnamefont {Kennedy}},
  \bibinfo {author} {\bibfnamefont {E.~H.}\ \bibnamefont {Lieb}}, \ and\
  \bibinfo {author} {\bibfnamefont {H.}~\bibnamefont {Tasaki}},\ }\href
  {\doibase 10.1103/PhysRevLett.59.799} {\bibfield  {journal} {\bibinfo
  {journal} {Phys. Rev. Lett.}\ }\textbf {\bibinfo {volume} {59}},\ \bibinfo
  {pages} {799} (\bibinfo {year} {1987})}\BibitemShut {NoStop}%
\bibitem [{\citenamefont {Affleck}\ \emph {et~al.}(1988)\citenamefont
  {Affleck}, \citenamefont {Kennedy}, \citenamefont {Lieb},\ and\ \citenamefont
  {Tasaki}}]{AKLT88}%
  \BibitemOpen
  \bibfield  {author} {\bibinfo {author} {\bibfnamefont {I.}~\bibnamefont
  {Affleck}}, \bibinfo {author} {\bibfnamefont {T.}~\bibnamefont {Kennedy}},
  \bibinfo {author} {\bibfnamefont {E.~H.}\ \bibnamefont {Lieb}}, \ and\
  \bibinfo {author} {\bibfnamefont {H.}~\bibnamefont {Tasaki}},\ }\href
  {\doibase 10.1007/BF01218021} {\bibfield  {journal} {\bibinfo  {journal}
  {Communications in Mathematical Physics}\ }\textbf {\bibinfo {volume}
  {115}},\ \bibinfo {pages} {477} (\bibinfo {year} {1988})}\BibitemShut
  {NoStop}%
\bibitem [{\citenamefont {Kl{\"u}mper}\ \emph {et~al.}(1992)\citenamefont
  {Kl{\"u}mper}, \citenamefont {Schadschneider},\ and\ \citenamefont
  {Zittartz}}]{Klumper92}%
  \BibitemOpen
  \bibfield  {author} {\bibinfo {author} {\bibfnamefont {A.}~\bibnamefont
  {Kl{\"u}mper}}, \bibinfo {author} {\bibfnamefont {A.}~\bibnamefont
  {Schadschneider}}, \ and\ \bibinfo {author} {\bibfnamefont {J.}~\bibnamefont
  {Zittartz}},\ }\href {\doibase 10.1007/BF01309281} {\bibfield  {journal}
  {\bibinfo  {journal} {Zeitschrift f{\"u}r Physik B Condensed Matter}\
  }\textbf {\bibinfo {volume} {87}},\ \bibinfo {pages} {281} (\bibinfo {year}
  {1992})}\BibitemShut {NoStop}%
\bibitem [{\citenamefont {Haegeman}\ \emph {et~al.}(2011)\citenamefont
  {Haegeman}, \citenamefont {Cirac}, \citenamefont {Osborne}, \citenamefont
  {Pi\ifmmode~\check{z}\else \v{z}\fi{}orn}, \citenamefont {Verschelde},\ and\
  \citenamefont {Verstraete}}]{verstrate11}%
  \BibitemOpen
  \bibfield  {author} {\bibinfo {author} {\bibfnamefont {J.}~\bibnamefont
  {Haegeman}}, \bibinfo {author} {\bibfnamefont {J.~I.}\ \bibnamefont {Cirac}},
  \bibinfo {author} {\bibfnamefont {T.~J.}\ \bibnamefont {Osborne}}, \bibinfo
  {author} {\bibfnamefont {I.}~\bibnamefont {Pi\ifmmode~\check{z}\else
  \v{z}\fi{}orn}}, \bibinfo {author} {\bibfnamefont {H.}~\bibnamefont
  {Verschelde}}, \ and\ \bibinfo {author} {\bibfnamefont {F.}~\bibnamefont
  {Verstraete}},\ }\href {\doibase 10.1103/PhysRevLett.107.070601} {\bibfield
  {journal} {\bibinfo  {journal} {Phys. Rev. Lett.}\ }\textbf {\bibinfo
  {volume} {107}},\ \bibinfo {pages} {070601} (\bibinfo {year}
  {2011})}\BibitemShut {NoStop}%
\bibitem [{\citenamefont {Haegeman}\ \emph
  {et~al.}(2013{\natexlab{a}})\citenamefont {Haegeman}, \citenamefont
  {Osborne},\ and\ \citenamefont {Verstraete}}]{Vers13}%
  \BibitemOpen
  \bibfield  {author} {\bibinfo {author} {\bibfnamefont {J.}~\bibnamefont
  {Haegeman}}, \bibinfo {author} {\bibfnamefont {T.~J.}\ \bibnamefont
  {Osborne}}, \ and\ \bibinfo {author} {\bibfnamefont {F.}~\bibnamefont
  {Verstraete}},\ }\href {\doibase 10.1103/PhysRevB.88.075133} {\bibfield
  {journal} {\bibinfo  {journal} {Phys. Rev. B}\ }\textbf {\bibinfo {volume}
  {88}},\ \bibinfo {pages} {075133} (\bibinfo {year}
  {2013}{\natexlab{a}})}\BibitemShut {NoStop}%
\bibitem [{\citenamefont {Haegeman}\ \emph
  {et~al.}(2013{\natexlab{b}})\citenamefont {Haegeman}, \citenamefont
  {Michalakis}, \citenamefont {Nachtergaele}, \citenamefont {Osborne},
  \citenamefont {Schuch},\ and\ \citenamefont {Verstraete}}]{verstraete13}%
  \BibitemOpen
  \bibfield  {author} {\bibinfo {author} {\bibfnamefont {J.}~\bibnamefont
  {Haegeman}}, \bibinfo {author} {\bibfnamefont {S.}~\bibnamefont
  {Michalakis}}, \bibinfo {author} {\bibfnamefont {B.}~\bibnamefont
  {Nachtergaele}}, \bibinfo {author} {\bibfnamefont {T.~J.}\ \bibnamefont
  {Osborne}}, \bibinfo {author} {\bibfnamefont {N.}~\bibnamefont {Schuch}}, \
  and\ \bibinfo {author} {\bibfnamefont {F.}~\bibnamefont {Verstraete}},\
  }\href {\doibase 10.1103/PhysRevLett.111.080401} {\bibfield  {journal}
  {\bibinfo  {journal} {Phys. Rev. Lett.}\ }\textbf {\bibinfo {volume} {111}},\
  \bibinfo {pages} {080401} (\bibinfo {year} {2013}{\natexlab{b}})}\BibitemShut
  {NoStop}%
\bibitem [{\citenamefont {Kim}\ \emph {et~al.}(2019)\citenamefont {Kim},
  \citenamefont {Pal},\ and\ \citenamefont {Han}}]{KPH19}%
  \BibitemOpen
  \bibfield  {author} {\bibinfo {author} {\bibfnamefont {J.}~\bibnamefont
  {Kim}}, \bibinfo {author} {\bibfnamefont {R.}~\bibnamefont {Pal}}, \ and\
  \bibinfo {author} {\bibfnamefont {J.~H.}\ \bibnamefont {Han}},\ }\href
  {\doibase 10.1103/PhysRevB.100.155104} {\bibfield  {journal} {\bibinfo
  {journal} {Phys. Rev. B}\ }\textbf {\bibinfo {volume} {100}},\ \bibinfo
  {pages} {155104} (\bibinfo {year} {2019})}\BibitemShut {NoStop}%
\bibitem [{\citenamefont {{Uimin}}(1970)}]{Uimin70}%
  \BibitemOpen
  \bibfield  {author} {\bibinfo {author} {\bibfnamefont {G.~V.}\ \bibnamefont
  {{Uimin}}},\ }\href@noop {} {\bibfield  {journal} {\bibinfo  {journal} {ZhETF
  Pisma Redaktsiiu}\ }\textbf {\bibinfo {volume} {12}},\ \bibinfo {pages} {332}
  (\bibinfo {year} {1970})}\BibitemShut {NoStop}%
\bibitem [{\citenamefont {Lai}(1974)}]{Lai74}%
  \BibitemOpen
  \bibfield  {author} {\bibinfo {author} {\bibfnamefont {C.~K.}\ \bibnamefont
  {Lai}},\ }\href {\doibase 10.1063/1.1666522} {\bibfield  {journal} {\bibinfo
  {journal} {Journal of Mathematical Physics}\ }\textbf {\bibinfo {volume}
  {15}},\ \bibinfo {pages} {1675} (\bibinfo {year} {1974})},\ \Eprint
  {http://arxiv.org/abs/https://doi.org/10.1063/1.1666522}
  {https://doi.org/10.1063/1.1666522} \BibitemShut {NoStop}%
\bibitem [{\citenamefont {Sutherland}(1975)}]{Sutherland75}%
  \BibitemOpen
  \bibfield  {author} {\bibinfo {author} {\bibfnamefont {B.}~\bibnamefont
  {Sutherland}},\ }\href {\doibase 10.1103/PhysRevB.12.3795} {\bibfield
  {journal} {\bibinfo  {journal} {Phys. Rev. B}\ }\textbf {\bibinfo {volume}
  {12}},\ \bibinfo {pages} {3795} (\bibinfo {year} {1975})}\BibitemShut
  {NoStop}%
\bibitem [{\citenamefont {Lozano}\ \emph {et~al.}(2013)\citenamefont {Lozano},
  \citenamefont {Murugan},\ and\ \citenamefont {Prinsloo}}]{CP3}%
  \BibitemOpen
  \bibfield  {author} {\bibinfo {author} {\bibfnamefont {Y.}~\bibnamefont
  {Lozano}}, \bibinfo {author} {\bibfnamefont {J.}~\bibnamefont {Murugan}}, \
  and\ \bibinfo {author} {\bibfnamefont {A.}~\bibnamefont {Prinsloo}},\ }\href
  {\doibase 10.1007/JHEP08(2013)109} {\bibfield  {journal} {\bibinfo  {journal}
  {Journal of High Energy Physics}\ }\textbf {\bibinfo {volume} {2013}},\
  \bibinfo {pages} {109} (\bibinfo {year} {2013})}\BibitemShut {NoStop}%
\bibitem [{\citenamefont {Arovas}\ \emph {et~al.}(1988)\citenamefont {Arovas},
  \citenamefont {Auerbach},\ and\ \citenamefont {Haldane}}]{AAH88}%
  \BibitemOpen
  \bibfield  {author} {\bibinfo {author} {\bibfnamefont {D.~P.}\ \bibnamefont
  {Arovas}}, \bibinfo {author} {\bibfnamefont {A.}~\bibnamefont {Auerbach}}, \
  and\ \bibinfo {author} {\bibfnamefont {F.~D.~M.}\ \bibnamefont {Haldane}},\
  }\href {\doibase 10.1103/PhysRevLett.60.531} {\bibfield  {journal} {\bibinfo
  {journal} {Phys. Rev. Lett.}\ }\textbf {\bibinfo {volume} {60}},\ \bibinfo
  {pages} {531} (\bibinfo {year} {1988})}\BibitemShut {NoStop}%
\bibitem [{\citenamefont {Green}\ \emph {et~al.}(2016)\citenamefont {Green},
  \citenamefont {Hooley}, \citenamefont {Keeling},\ and\ \citenamefont
  {Simon}}]{green}%
  \BibitemOpen
  \bibfield  {author} {\bibinfo {author} {\bibfnamefont {A.~G.}\ \bibnamefont
  {Green}}, \bibinfo {author} {\bibfnamefont {C.~A.}\ \bibnamefont {Hooley}},
  \bibinfo {author} {\bibfnamefont {J.}~\bibnamefont {Keeling}}, \ and\
  \bibinfo {author} {\bibfnamefont {S.~H.}\ \bibnamefont {Simon}},\ }\href@noop
  {} {\enquote {\bibinfo {title} {Feynman path integrals over entangled
  states},}\ } (\bibinfo {year} {2016}),\ \Eprint
  {http://arxiv.org/abs/1607.01778} {arXiv:1607.01778 [cond-mat.str-el]}
  \BibitemShut {NoStop}%
\bibitem [{\citenamefont {Hallam}(2019)}]{hallam}%
  \BibitemOpen
  \bibfield  {author} {\bibinfo {author} {\bibfnamefont {A.}~\bibnamefont
  {Hallam}},\ }\emph {\bibinfo {title} {Tensor network descriptions of quantum
  entanglement in path integrals, thermalisation and machine learning}},\
  \href@noop {} {Ph.D. thesis},\ \bibinfo  {school} {UCL (University College
  London)} (\bibinfo {year} {2019})\BibitemShut {NoStop}%
\bibitem [{\citenamefont {Seifie}\ \emph {et~al.}(2019)\citenamefont {Seifie},
  \citenamefont {Singh},\ and\ \citenamefont {Mathey}}]{mathey}%
  \BibitemOpen
  \bibfield  {author} {\bibinfo {author} {\bibfnamefont {I.~M.~H.}\
  \bibnamefont {Seifie}}, \bibinfo {author} {\bibfnamefont {V.~P.}\
  \bibnamefont {Singh}}, \ and\ \bibinfo {author} {\bibfnamefont
  {L.}~\bibnamefont {Mathey}},\ }\href {\doibase 10.1103/PhysRevA.100.013602}
  {\bibfield  {journal} {\bibinfo  {journal} {Phys. Rev. A}\ }\textbf {\bibinfo
  {volume} {100}},\ \bibinfo {pages} {013602} (\bibinfo {year}
  {2019})}\BibitemShut {NoStop}%
\end{thebibliography}%
\end{document}